%% file: Main.tex
\documentclass[acmsmall,natbib=false]{acmart}
\pdfoutput=1
\usepackage{graphicx}
\usepackage[export]{adjustbox}
\usepackage{subcaption}
\usepackage{enumitem}
\usepackage{multirow, makecell}
\newsavebox{\measurebox}
 \usepackage{wrapfig}

\setcopyright{none}
 \renewcommand\footnotetextcopyrightpermission[1]{}
\usepackage[backend=bibtex,
style=ACM-Reference-Format,
bibencoding=ascii,
sorting=ynt]{biblatex}
\begin{document}

\title{A Taxonomy of Knowledge Gaps for Wikimedia Projects (Second Draft)}

\author{Miriam Redi}
\affiliation{%
  \institution{Wikimedia Foundation}
}
\author{Martin Gerlach}
\affiliation{%
  \institution{Wikimedia Foundation}
}
\author{Isaac Johnson}
\affiliation{%
  \institution{Wikimedia Foundation}
}
\author{Jonathan Morgan}
\affiliation{%
  \institution{Wikimedia Foundation}
}
\author{Leila Zia}
\affiliation{%
  \institution{Wikimedia Foundation}
}

\maketitle
\clearpage
\input{sections/0_Executive_Summary}

\clearpage
\tableofcontents
\clearpage
\input{sections/1_Introduction}
\input{sections/2_Related}
\input{sections/3_Method}
\input{sections/4_a_Readers_Taxonomy}
\input{sections/4_b_Contributors_Taxonomy}
\input{sections/4_c_Content_Taxonomy}
\input{sections/5_How_to}
\input{sections/6_Future}
\input{sections/7_Acknowledgements}

\printbibliography[keyword={literature}, title={Academic Literature}]
\printbibliography[keyword={projects}, title={Movement Projects}]
\printbibliography[keyword={community initiatives}, title={Community Initiatives and Events}]
\printbibliography[keyword={surveys}, title={Surveys}]
\printbibliography[keyword={strategy}, title={Strategic Directions}]
\printbibliography[keyword={policies}, title={Policies, Guidelines and Discussions}]
\printbibliography[keyword={tools}, title={Tools}]

\end{document}

%% file: sections/0_Executive_Summary.tex
\section*{Executive Summary}
In January 2019, prompted by the Wikimedia Movement's 2030 strategic direction~\cite{strategy}, the Research team at the Wikimedia Foundation\footnote{\url{https://research.wikimedia.org/team.html}} identified the need to develop a \emph{knowledge gaps index}---a composite index to support the decision makers across the Wikimedia movement by providing: a framework to encourage structured and targeted brainstorming discussions; data on the state of the knowledge gaps across the Wikimedia projects that can inform decision making and assist with measuring the long term impact of large scale initiatives in the Movement. 

After its first release in July 2020, the Research team has developed the second complete draft of a taxonomy of knowledge gaps for the Wikimedia projects, as the first step towards building the knowledge gap index. We studied more than 250 references by scholars, researchers, practitioners, community members and affiliates---exposing evidence of knowledge gaps in readership, contributorship, and content of Wikimedia projects. We elaborated the findings and compiled the taxonomy of knowledge gaps in this paper, where we describe, group and classify knowledge gaps into a structured framework. The taxonomy that you will learn more about in the rest of this work will serve as a basis to operationalize and quantify knowledge equity, one of the two 2030 strategic directions, through the knowledge gaps index. 

We hope you enjoy the read and join the conversation about how we can improve the taxonomy of knowledge gaps for the Wikimedia projects. If you have any suggestion or feedback, please reach out to us via the Knowledge Gaps Index Meta page.\footnote{\url{https://meta.wikimedia.org/wiki/Research:Knowledge_Gaps_Index/Taxonomy}}

%% file: sections/1_Introduction.tex
\section{Introduction}

%% file: sections/2_Related.tex
With almost 56 million articles written by roughly 500,000 monthly editors across more than 160 actively edited languages, Wikipedia is the most important source of encyclopedic knowledge and one of the most important knowledge resources available on the internet. Every month, the project attracts users on more than 1.5 billion unique devices from across the globe, for a total of more than 15 billion monthly pageviews~\cite{stats2}.

While global and massively popular resources, Wikipedia and its sister projects such as Wikimedia Commons and Wikidata suffer from a wide range of \textit{knowledge gaps}, which we define as \textit{disparities in content coverage or participation of a specific group of readers or contributors.}

A typical example of a knowledge gap is the gender gap, one of the most well-studied gaps in the Wikiverse. Researchers and practitioners have investigated the gender gap by measuring the representation of different gender groups in content, readers, and contributors, and found, for example, that less than 20\% of the biographies in Wikipedia are about women~\cite{whgi}, roughly 75\% of readers in Wikipedia are men~\cite{johnson2020global}, and that this disparity becomes even more extreme when analyzing editors' gender distribution~\cite{2020communityinsights}.

Beyond the gender gap, only a handful of works have studied ways to quantify or address other kinds of knowledge gaps---for example by analyzing readers'~\cite{lemmerich2019world} and editors'~\cite{cho2010testing} motivations for accessing the site, studying the role of technical skills and awareness in Wikimedia contributorship~\cite{menking2015heart,hargittai2015mind} and readership~\cite{shaw2018pipeline}, designing algorithms to identify and prioritize missing content in different languages~\cite{wulczyn2016growing}, or proposing question-and-answer facilities to satisfy readers' information needs~\cite{chhabra2016should}.

While these works made substantial progress towards measuring and identifying directions for bridging individual gaps, they did not provide a holistic and systematic framework to understand the scope of knowledge gaps in Wikimedia and the relationships between them. 

To see why having such a comprehensive map of the space of knowledge gaps could be key for Wikimedia projects, consider the main directions identified as part of the Movement Strategy~\cite{strategy}: by 2030, Wikimedia will become the essential infrastructure of the ecosystem of free knowledge, and achieve \textit{knowledge equity} by focusing efforts on including all ``knowledge and communities that have been left out by structures of power and privilege''. To support this goal, and measure the progress towards knowledge equity, the Research team at the Wikimedia Foundation has identified the need to operationalize knowledge equity into a \emph{knowledge gaps index} -- a composite index tracking the collective evolution of knowledge gaps in Wikimedia\footnote{While it is hard to make the claim that understanding gaps is the most effective way to understand inequalities, what we observe is that understanding gaps is a way to spark conversations and motivate/mobilize towards reaching knowledge equity. However, we acknowledge that inequalities are usually highly complex and the abstraction level that gaps introduce can over-simplify the deep understandings one can gain through ethnographic research and focus too much on deficits and not enough on the success of various initiatives, for example.}.

While extremely important, operationalizing knowledge equity by studying and measuring its underlying factors is not a trivial task. The first step towards this goal is to generate a systematic picture of those Wikimedia audiences, groups, and cultures that could be  underrepresented in terms of participation, representation, and coverage. 
   
\begin{figure*}[!h]
    \centering
    \includegraphics[width=0.8\linewidth]{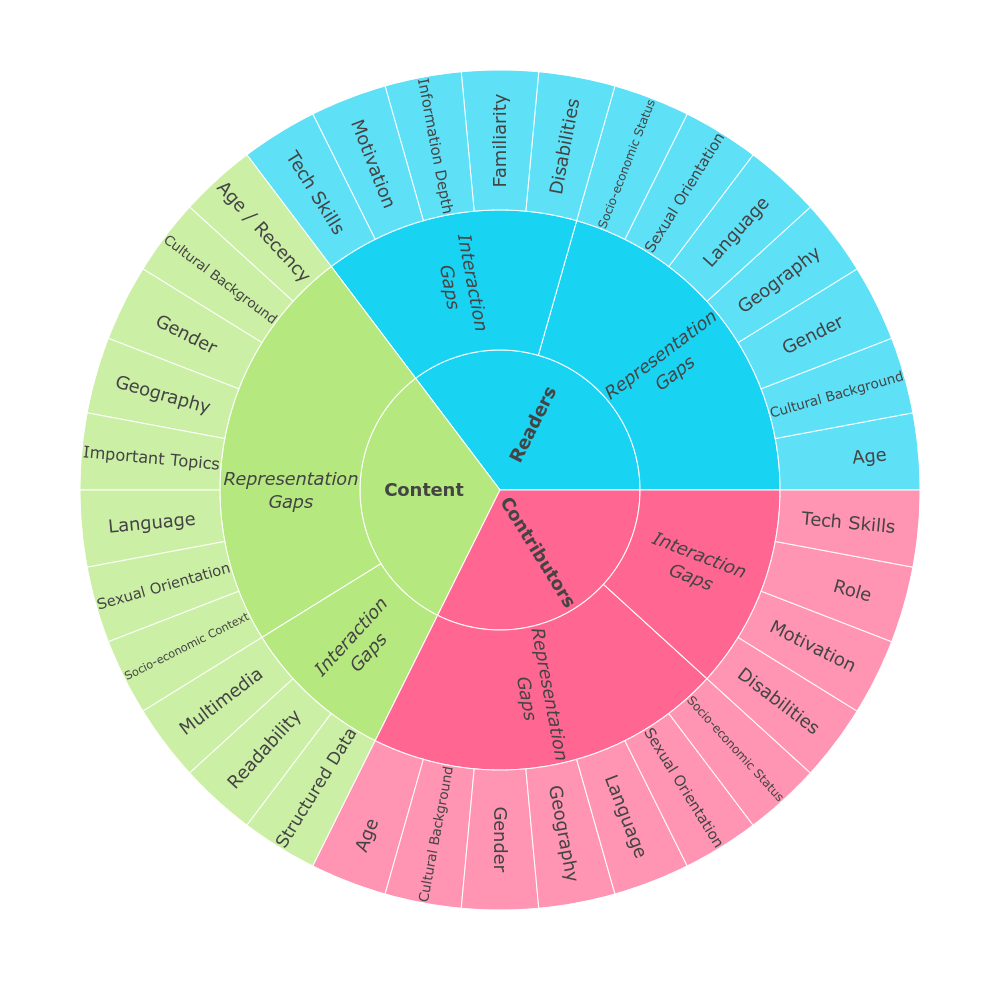}
    \vspace{-1cm}
    \caption{The Knowledge Gaps Taxonomy}
    \label{fig:taxonomy_wheel}
\end{figure*}

Therefore, in this work we propose the first \textit{taxonomy of knowledge gaps}\footnote{Following strict definitions~\cite{smith2002typologies}, the final product should be defined as a \textit{typology} rather than a \textit{taxonomy}, as ``[t]axonomies differ from typologies in that they classify items on the basis of empirically observable and measurable characteristics''~\cite{smith2002typologies}, while typologies group concepts according to characteristics which are less tangible and might not exist in physical reality. However, given the widespread usage of the term ``taxonomy'' to classify objects according to abstract properties---e.g., emotions, clustering algorithms, or educational objectives~\cite{tamir2016people,fahad2014survey,bloom1956taxonomy}---we will use this term to define the final product of our work throughout the manuscript.} in the context of Wikimedia projects. We identify three macro-dimensions of the Wikimedia ecosystem as the roots of the taxonomy: Readers, Contributors, and Content. We then review studies and discussions from scholars, researchers, practitioners and community members, and compile for each of these dimensions a list of knowledge gaps as areas of the Wikiverse where we found systematic evidence of inequality, together with potential \textit{barriers} which prevent us from addressing these knowledge gaps. Finally, we group these gaps into \textit{facets} containing knowledge gaps related to \textit{Representation} or \textit{Interaction} aspects. The final three-layer taxonomy is illustrated in Figure~\ref{fig:taxonomy_wheel}.
   
This taxonomy provides a theoretical framework based on which we can develop different approaches to address knowledge gaps and represents a first step towards the full operationalization of knowledge equity. With this taxonomy, we aim at fostering conversations across the Wikimedia and academic communities around the nature and composition of the ``content and communities that have been left out by structures of power and privilege''~\cite{strategy}. Future work will include the development of robust metrics and indicators to quantify gaps, which could allow to track progress towards knowledge equity~\cite{Commission2008handbook}. Quantifying gaps will facilitate the conceptualization of the interdependence between different gaps in order to understand potential causes and barriers, and to design more effective interventions. 
   
In the remainder of this paper, we will first look at the methodology used to compile the taxonomy (Section~\ref{sec:methods}), then present each macro branch of the taxonomy: Readers (Section~\ref{sec:readers}), Contributors (Section~\ref{sec:contributors}), and Content (Section~\ref{sec:content}). We will then provide some suggestions on how different audiences can use and benefit from this taxonomy (Section~\ref{sec:how-to}), and finally expose interesting directions for future work (Section~\ref{sec:future}).

%% file: sections/3_Method.tex
\section{Methodology}\label{sec:methods}
The Taxonomy of Knowledge Gaps emerges as the result of hours of literature review, survey analysis, in-depth movement strategy reading, research brainstorming and Wikimedia Foundation-wide discussions. To help understand the methodology we used to compile this taxonomy, in this Section we explain its primary guiding principles and structural characteristics.
\subsection{Guiding Principles}\label{sec:guidingprinciples}
Our task is to collect elements of Wikimedia spaces lacking knowledge equity in a taxonomy of knowledge gaps. There are endless ways in which we could bring order to this large pool of unstructured pieces of information. To help in narrowing down the scope while delivering a consistent final product, we let different themes, principles and values guide the selection of the taxonomy classes and levels. 
\begin{description}
    \item[Driven by communities.] The taxonomy is largely inspired by values and principles of the Wikimedia communities and the latest recommendations of the Movement Strategy~\cite{strategy}. At the same time, the taxonomy is grounded in the literature produced by the broader academic and scientific communities who study Wikimedia space from a computational and sociological perspective (for an overview, see reviews~\cite{Okoli2012review,Nielsen2012review,Mesgari2015sum} or the Wikimedia Research Newsletter~\cite{wrn}).
    \item[Neutrality.] Similar to Bloom's taxonomy of educational objectives~\cite{krathwohl2009taxonomy}, one of the foundational principles of this taxonomy is its neutrality. While defining gaps is an inherently value-laden process, our mission is to define a classification of knowledge gaps, and identify some of the barriers preventing people from accessing free knowledge. While doing so, we aim at being as impartial and inclusive as possible, without expressing value judgments on the importance of one gap or source over another.
    \item[Flexibility.] Like any project in Wikimedia spaces, we wanted this taxonomy to be ``editable''. This is the second version of the taxonomy of knowledge gaps. We made the first version of the taxonomy available to the community and opened it up for feedback from researchers and community members. After one month of consultation, we worked on summarizing and incorporating the precious suggestions we received \footnote{For a summary of changes, see here: \url{https://meta.wikimedia.org/wiki/Research:Knowledge_Gaps_Index/Taxonomy/Summary_of_Changes_for_Second_Version}}, which are now embedded in this second version. Also, its inherent structure allows the taxonomy to evolve and be expanded by experts and community members who want to contribute to the improvement of its content.
    \item[Measurability.] One of the end goals of this taxonomy is to help Wikimedia communities measure the impact of their initiatives and content creation. With this in mind, when possible, we explicitly formulated gaps and objectives to incorporate elements that are quantifiable in a globally-consistent manner via surveys, large-scale data analysis or other computational methods.
\end{description}

\subsection{Sources}
To understand relevant components of the taxonomy of knowledge gaps and make decisions about which aspects of Wikimedia ecosystems to include in the taxonomy, we gathered information from different sources describing or discussing inequalities among readers, contributors, and content in Wikimedia spaces. We explain here our main data sources and the rationale behind choosing each of them.
\begin{description}
    \item[Academic Literature.] We tap into ideas from several fields of academic literature that study knowledge gaps in Wikimedia. The majority of this research belongs to the broad field of ``computational social science'' and tries to characterize and quantify different aspects of Wikimedia communities using a computational approach. For example, researchers have studied gender gaps across different dimensions~\cite{Hinnosaar2019newmedia,konieczny2018gender}, quantified the usage of visual content across different languages~\cite{He2018babel} or estimated the geographical bias of Wikipedia content~\cite{Beytia2020positioning}. Research aiming to bridge knowledge gaps using tools from recommender systems and natural language processing also largely helped the creation of this taxonomy: these include methods to grow Wikipedia languages via recommendations~\cite{wulczyn2016growing}, as well as machine learning models to score readability of Wikipedia articles~\cite{brezar12019readability}, or discover sentences needing citations~\cite{Redi2019citationneeded}.
    \item[Community Surveys.] Throughout the years, the Wikimedia Foundation, Wikimedia Affiliates, and independent organizations have run surveys to characterize Wikimedia readers and editors. 
    We tap into the questions asked in these surveys to define some of the gaps and facets in our taxonomy, for example many of our representational gaps. Most of these surveys are referenced throughout this manuscript, and a subset of them can be found by checking the appropriate categories on \url{meta.wikimedia.org}.\footnote{\url{https://meta.wikimedia.org/wiki/Category:Surveys}; \url{https://meta.wikimedia.org/wiki/Category:Reader_surveys}; \url{https://meta.wikimedia.org/wiki/Category:Editor_surveys}}
    \item[Movement Strategy.] The third source of content for this taxonomy is a set of guidelines from the Wikimedia Movement Strategy. Such guidelines include broad strategic directions, namely Knowledge Equity and Knowledge as a Service~\cite{strategy}, as well as more specific recommendations to implement such strategic directions~\cite{uxstrategy,innovationstrategy}. We also borrow from the Wikimedia Foundation's medium-term-plan priorities~\cite{mtp}. Collectively, these guidelines helped us to identify gaps, barriers, and common objectives related to free knowledge.
    \item[Community Initiatives.] Throughout the years, the Wikimedia community have worked on many initiatives and discussions aiming to address knowledge gaps. These not only include research (e.g. trying to conceptualize or measure the gap) but also cover a wide variety of initiatives such as WikiProjects (e.g,. WikiProject Women~\cite{wikiprojectwomen}) or campaigns (e.g., Wiki Loves Monuments~\cite{wikilovesmonuments}).
\end{description}
\subsection{Taxonomy Structure}
Beyond the identification of potential gaps within the Wikimedia ecosystem, one of the main challenges in building a taxonomy of knowledge gaps is to provide a structure for how to analyze and classify the different gaps.
In this section we give an overview over the overall structure, and define and motivate the different levels in the taxonomy.

\begin{description}
    \item[Gap.] A \textit{gap} corresponds to an individual aspect of the Wikimedia ecosystem---for example readers' gender, or images in content---for which we found evidence of a lack of diversity, or imbalanced coverage across its inner categories (for example, proportion of readers who identify as men, women or non-binary in the case of the reader gender gap). 
    For each gap, we include the following characterizing fields:
    \begin{itemize}
    \item \textit{Description}: This field is the definition of the gap, and describes which type of disparity is covered by the gap. For example, the reader gender gap is defined as ``the difference between readers of different genders in how and how much they access the sites''.
    \item \textit{Sources}: The decision to include a gap in the taxonomy is heavily driven by existing sources. The source property of a gap includes all the references we examined when characterising and describing a given gap, grouped by source type (literature, surveys, strategy and/or community).
    \end{itemize}

    \item[Facet.] We grouped semantically-related gaps into \textit{facets}, namely macro-categories describing the general semantics of a gap. For example, gaps such as ``gender'' or ``age'' both belong to the ``Representation'' facet.
    
    \item[Dimension.] 
    We identify three foundational \textit{dimensions} as root nodes in our taxonomy grouping different facets: Wikimedia readers, Wikimedia contributors, and Wikimedia content. We will expand the rationale behind the choice of the dimensions in the next Subsection.
    
    \item[Barriers.] Defining an exhaustive list of gaps' potential causes---i.e. the technical, social, and economic barriers preventing the Wikimedia ecosystem from closing knowledge gaps and achieving greater diversity---lie beyond the scope of this manuscript. Still, throughout our literature review of knowledge gaps in Wikipedia and sister projects, we found evidence of elements which could potentially amplify or cause inequalities, and describe them, for each dimension, in a \textit{Barriers} section. 
\end{description}

\subsection{Root Dimensions: Readers, Contributors, and Content}
    This taxonomy is developed along three dimensions representing macro-elements of the Wikimedia ecosystem: Readers, Contributors, and Content. The choice of these dimensions is broadly inspired by two models summarizing the inner mechanisms of Wikimedia communities. 
    
\begin{figure}
\begin{subfigure}{.48\linewidth}
    \centering
    \includegraphics[width=.8\linewidth]{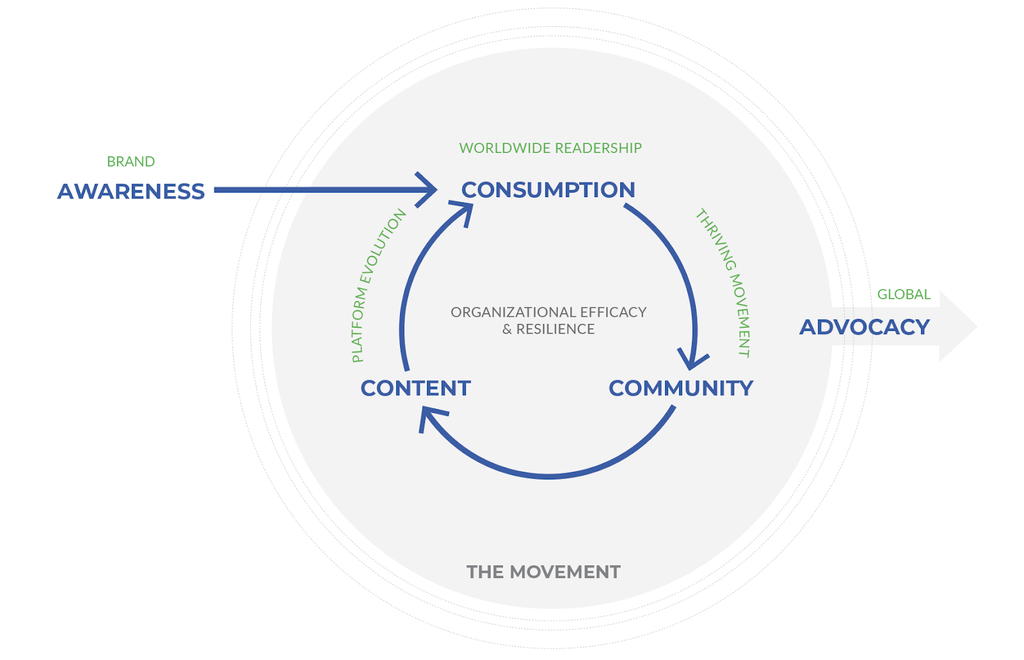}
    \caption{Margeigh Novotny / CC BY-SA}
    \label{fig:mtp_model}
\end{subfigure}
\begin{subfigure}{.48\linewidth}
    \centering
    \includegraphics[width=.8\linewidth]{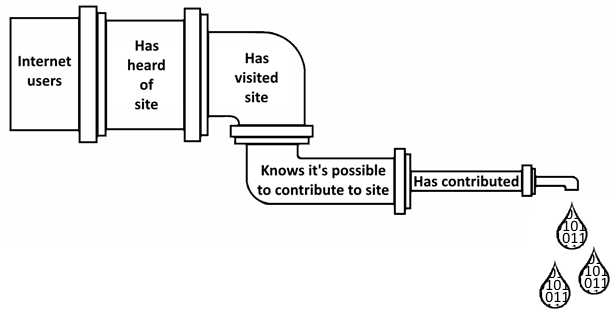}
    \caption{Aaronshaw / CC BY-SA}
    \label{fig:pipeline_model}
\end{subfigure}
\caption{Sketch of the two models on the complex dynamics in  Wikimedia communities giving rise to the three root dimensions (readers, contributors, content) in the taxonomy of knowledge gaps.
}
\label{fig:models}
\end{figure}

    A model proposed by Wikimedia Foundation as part of their medium-term plan~\cite{mtp} for how to build engagement across the Movement sketches a diagram in the form of a flywheel describing the relationship between awareness, consumption, contributors, content, and advocacy, as shown in Figure~\ref{fig:mtp_model}.  
    The model shows how people initially engage with the site (i.e. they become Wikimedia \textbf{readers}) through spontaneous search or awareness campaigns. Some readers then might become more involved in the values or the mission of the Wikimedia movement, thus potentially converting into \textbf{contributors} or donors. When more people engage with the site, the \textbf{content} production becomes greater and more diverse, thus making the site more inclusive for new readers, and the cycle begins again. 
    
    A complementary model by Shaw and Hargittai~\cite{shaw2018pipeline} (``The Pipeline of Online Participation Inequalities: The Case of Wikipedia Editing'') proposes a framework in the form of a pipeline to describe how gaps form at different levels of engagement with Wikipedia (see Figure~\ref{fig:pipeline_model}). Specifically, they propose a similar series of steps through which a user must go to become a contributor: being an internet user, having heard of the site (awareness of Wikipedia), having visited the site (Wikipedia \textbf{reader}), knowing it's possible to contribute to the site (awareness they can edit), and then finally having contributed (is a \textbf{contributor}). At each of these steps, they identify structural barriers to participation that make different groups of people more or less likely to drop off in participation. As a result, gaps that were perhaps non-existent or small at the earlier step of awareness become much larger at the later step of contribution.
    
    The research cited within this taxonomy bears out the interconnectedness of reader, contributor, and content gaps as described by these models. Together, these two models demonstrate both the importance of initiatives that seek to expand the diversity of the reader population, contributor population, and content, \textit{and}, the need to address barriers to awareness and participation at each step of joining the Wikimedia Movement.

For each dimension of this Knowledge Gaps Taxonomy---\textit{Readership} (Section~\ref{sec:readers}), \textit{Contributorship} (Section~\ref{sec:contributors}) and,  \textit{Content} (Section~\ref{sec:content})---we present a short overview of its meaning and the corresponding taxonomy in a tabular form. For each of the dimension's facet, we provide a detailed explanation for the rationale behind it.

%% file: sections/4_a_Readers_Taxonomy.tex
\input{tables/readership_table}
\section{Readers}
\label{sec:readers}
The readership dimension of knowledge gaps encompasses all those gaps related to readers' access to Wikimedia sites. We identify different areas where readers are under-represented according to literature, surveys, and community strategic directions and organize them around two main facets: representation gaps and interaction gaps. We also highlight different initiatives aiming at closing each gap. We define readers as all users who connect to the site to \textit{consume} Wikimedia content. While there exists a body of research~\cite{alshomary2019wikipedia,mcmahon2017substantial,rothshild2019interplay,vincent2019measuring,vincent2018examining,erickson2018commons} studying how content consumption happens outside of Wikimedia---e.g., voice assistants, search engines, or third-party apps---we scope down our definition of readership to readers who come directly to the projects to access content.

A large source of inspiration for this dimension is the 2020 Wikimedia Movement Strategy, which recommends that Wikimedia platforms should be designed to ``enable everyone---irrespective of gender, culture, age, technological background or skills, or physical abilities---to enjoy a positive experience during both consumption and contribution to knowledge throughout the Wikimedia ecosystem''~\cite{uxstrategy}.
\subsection{Representation Gaps}
Representation readership gaps have been widely discussed across the Wikimedia movement, exposed by surveys organized by the Wikimedia Foundation and the chapters, and analyzed by research studying the Wikimedia ecosystem. This facet includes gaps in readership related to sociodemographics and identity such as gender, age, language and location.
\subsubsection{Gender}
The \textit{gender} gap is the difference between readers of different gender identities in how and how much they access the sites. 
The gender gap has been measured through many readership surveys organized by the Wikimedia Foundation~\cite{2011survey,2011phonesurvey,2014globalsouthsurvey,2015donationsurvey,2016phonesurvey,2017strategysurvey,2019survey}, different chapters~\cite{2012banglasurvey,2014pakistansurvey,2016welshsurvey,2013dutchsurvey}, independent organizations~\cite{2007pewsurvey,2010pewsurvey,2008survey} and research communities~\cite{shaw2018pipeline,protonotarios2016similar,Hinnosaar2019newmedia} by asking respondents for their gender identity.
This makes the gender gap one of the most heavily studied constructs with respect to readership. However, almost all surveys have only included men and women as gender identities and have not distinguished between cis and trans identities, making it difficult to e.g., draw any conclusions about readership among people with non-binary gender identities. 
Furthermore, conclusions about gender gaps might differ depending on the definition of readers. For example, the most recent survey in this space~\cite{2019survey} defined readers as people who read Wikipedia daily, and found that, in almost every country/language surveyed (with the exception of Romanian), there is a substantial gender gap---i.e. men are more likely to be frequent readers and tend to generate many more page views than other gender identities even in regions where previous surveys have shown no strong gender gaps in occasional usage. 
A number of factors cause some gender identities to be under-represented in Wikimedia's readership population, and across the years, different community initiatives have focused on bridging this gap~\cite{gendergapinitiatives}. 
\subsubsection{Age}
The \textit{age} gap reflects how and how much readers of different age access Wikimedia sites. There exists a large volume of data on readers' age, collected by surveys and academic literature~\cite{2011survey,2011phonesurvey,2014globalsouthsurvey,2015donationsurvey,2016phonesurvey,2017strategysurvey,2019survey,2012banglasurvey,2014pakistansurvey,2016welshsurvey,2007pewsurvey,2010pewsurvey,2008survey,shaw2018pipeline,protonotarios2016similar,Hinnosaar2019newmedia,2013dutchsurvey}. In cases where the age-distribution at country level is available, the data indicates that readers tend to be much younger than the general populace.  Among other things, the quality and presentation of content might affect this gap, and various initiatives across the movement focus on making content more available and readable for specific age ranges, such as Simple English Wikipedia~\cite{simplewikiabout}, the Wikijuniors project from Wikibooks~\cite{wikibooksjunior}, or WikiProject Accessibility~\cite{wikiprojectaccess}, which, among other things, aims at making content more accessible to elderly readers. 
\subsubsection{Geography}
The \textit{geography} gap refers to the different levels of readership across different geographic regions. This gap occurs primarily across two scales: between countries and within countries. For between countries, it is well-documented that there are varying levels of Wikipedia readership across different countries worldwide~\cite{wivivi,2016phonesurvey,2017strategysurvey,2011survey,2014globalsouthsurvey}. Many of the barriers discussed in Section~\ref{sec:readerbarriers} such as censorship or data cost are mediated by country-level policies and resources. Initiatives such as awareness campaigns~\cite{newreadersresearch} or Wikipedia Zero~\cite{wikipediazero} have been run in the past to boost readership in various countries.

The within-country aspect of geography largely refers to an individual's proximity to urban areas, which tends to be a strong proxy for availability of services. While there are no global standardized definitions for how to define urban/rural, which makes it difficult to compare self-reported locale to official statistics, results from Wikimedia-Foundation-led and academic surveys~\cite{2019survey,2016phonesurvey,shaw2018pipeline,2017strategysurvey} suggest strong over-representation of urban areas among the reader population. To help bridging this gap, projects such as WikiConnect~\cite{wikiconnect} and the Wikimedia Foundation's ``New Readers'' initiative~\cite{newreadersresearch} have worked on connecting Wikipedia with rural areas and those people who don't have direct access to Wikipedia.
\subsubsection{Language}
The \textit{language} gap reflects the different levels of readership depending on readers' ability to read one or more languages. What languages an individual can read greatly impacts what content is available to them and can introduce greater barriers if they are forced to read content in a language that is less familiar to them. 
Surveys have been conducted to estimate readers' literacy~\cite{2019survey,2016welshsurvey,2014globalsouthsurvey,2011phonesurvey} suggesting that certain languages have highly-literate readers.
For example, languages that are specific to one country show high levels of literacy amongst readers. 
In contrast, other languages such as English or French, which are more strongly associated with colonialism, have many readers for which English / French is not their native language~\cite{2019survey}. 
In order to address this issue, in English Simple Wikipedia was introduced using a simpler grammar and a limited vocabulary. While improving readability in comparison to English Wikipedia, research has shown that its level is still not ideal for readers with limited language literacy~\cite{lucassen2012readability}.
Other initiatives attempting to bridge this gap aim at making access to content in one's local language by growing under-represented languages such as Scribe~\cite{scribe}, the GapFinder tool~\cite{gapfinder}, Content Translation~\cite{contenttranslation}, or the Growing Local Language Content on Wikipedia initiative~\cite{glow}.

\subsubsection{Socioeconomic Status}
The \textit{socioeconomic status} gap is the difference in how readers with different education, income, wealth, or employment status access Wikimedia sites. All these aspects are highly correlated and difficult to disentangle. We start by looking at \textit{education}.
Education systems vary across different countries, and one commonly-used country-agnostic proxy for education level is the number of years of education. Surveys from researchers~\cite{shaw2018pipeline,Hinnosaar2019newmedia} and from the Wikimedia Foundation and affiliates~\cite{2019survey,2014pakistansurvey,2016welshsurvey,2014globalsouthsurvey,2008survey,2010pewsurvey,2010pewsurvey,2012banglasurvey,2011phonesurvey,2011survey,2013dutchsurvey} consistently demonstrate that individuals with higher levels of education are more likely to be readers of Wikipedia. Similar to education, \textit{income}, \textit{wealth}, and \textit{employment status} relate to an individual's means and class status.  Surveys~\cite{2007pewsurvey,2010pewsurvey,2011phonesurvey,2014pakistansurvey,2011survey,2012banglasurvey,2011phonesurvey,2014globalsouthsurvey} and recent research~\cite{shaw2018pipeline,Hinnosaar2019newmedia} consistently indicate that higher volumes of readership are found among people with higher income, though the relationship between employment status and readership is less clear. The other strong, and sometimes countervailing, trend is that students also tend to be much heavier readers of Wikipedia. Some Wikimedia projects are trying to close this gap, by specifically targeting readers with different levels of education, for example Wikiversity~\cite{wikiversity}; relatedly, initiatives such as ``Wikimedia+Education'' specifically focus on creating community around the topic of making Wikipedia more integrated with education systems at different levels~\cite{wikimediaeducation}.

\subsubsection{Sexual Orientation}
The \textit{sexual orientation} gap is the difference between readers who hold various sexual orientations and how they access the site. Cultural communities around sexual orientation often are tied to gender identity as well as with LGBTQ+ communities.\footnote{\url{https://meta.wikimedia.org/wiki/Wikimedia_LGBT\%2B/Portal}} We are unaware of empirical research that has specifically focused on sexual orientation on-wiki within the reader community but careful handling of sexual orientation on Wikipedia can be very important to readers who have these identities~\cite{wexelbaum2015queering}. Notably, given the oppression around sexual orientation in many countries around the world, extreme care should be taken before including questions about sexual orientation in surveys to readers.

\subsubsection{Cultural Background}
The \textit{cultural background} gap includes differences in readership among people with different ethnicity, race, political beliefs, or religious beliefs. These are gaps that tend to be very contextual and specific to regions---i.e. the relationship between any aspect of one's cultural background and the power they hold within society and representation on Wikimedia projects can vary greatly between regions. Studies on readers from emerging communities found some evidence that local socio-political constructs might impact the way in which Wikipedia's credibility is perceived~\cite{newreadersresearch}. Wikipedia by nature aims to structure its content to be welcoming to readers with different individual identities, perspectives and opinions. For example, the ``Neutral Point of View'' policy~\cite{npov} encourages editors to represent ``fairly, proportionately, and, as far as possible, without editorial bias, all the significant views that have been published by reliable sources on a topic''. Most friendly space policies in Wikimedia also encourage inclusiveness at many levels, including the respect of people's religious views. Moreover, surveys around community health report religion as one prominent factor creating barriers to involvement in Wikimedia communities~\cite{communityhealth}.

Ethnicity and race, or other proxies for caste, bring additional context to an individual's cultural background and often relate strongly with access to resources. Studies in the United States have shown gaps in usage along racial lines~\cite{2007pewsurvey,2010pewsurvey,shaw2018pipeline} and presumably these same trends are found in other countries in that people who are disenfranchised due to their race or ethnicity also would be less likely to access Wikipedia.
\subsection{Interaction Gaps}
This facet deals with all those gaps related to readers' ability and experience when interacting Wikimedia sites. Gaps in this facet include reader's motivation, familiarity with the content, and information need behind their presence on Wikipedia and sister projects, as well as individuals' abilities to interact with the sites. Notably, while readers' identities described under the Representation gaps are relatively fixed, Interaction gaps capture how a readers' changing context might impact access.

\subsubsection{Motivation}
The \textit{motivation} gaps reflect different levels of readership depending on the reason behind readers' visit to the site. There is quite a bit of variability between languages in readers' predominant motivations~\cite{lemmerich2019world,2017motivationsurvey}. The 2019 Reader Demographics Surveys~\cite{2019survey} confirmed that these differences cannot be fully explained by individual demographics.
\subsubsection{Information Depth}
The \textit{information depth} gaps reflect different levels of readership depending on what level of content a reader is looking for on the site---i.e. checking a quick fact, looking for an overview of a topic, or an in-depth read. Analogous to reader motivations, there is quite a bit of variability between languages in what readers' predominant information depths are. The 2019 Reader Demographics Surveys~\cite{2019survey} confirmed that these differences cannot be explained by individual demographics. Still not fully understood is how these information needs might relate to the utility of Wikipedia for a given reader and whether readers with information needs like ``fact'' or even ``overview'' are largely consuming Wikipedia via external re-use and are therefore underrepresented in the existing data. Additional information about reader information depth was gathered in past reader surveys~\cite{2019trustsurvey,lemmerich2019world,singer2017we}.
\subsubsection{Familiarity}
The \textit{familiarity} gaps reflect different levels of readership depending on one's prior familiarity with a topic. Similar to the motivation, someone's prior knowledge about a topic might affect their perception of Wikipedia as an appropriate data source, and average readers' prior knowledge on the topic they read varies across Wikipedia language editions~\cite{lemmerich2019world,2017motivationsurvey}.
Ongoing community discussions about content ``Depth vs Breadth'' indirectly reason about solutions to address these gaps~\cite{breadthdepth}.
\subsubsection{Tech Skills}
The \textit{tech skills} gap reflects different levels of readership depending on readers' general internet skills. This captures one's experience with the internet and ability to not only find the content one is looking for but also the ability to verify it.
Academic research~\cite{shaw2018pipeline} has found that high internet skills are associated with an increase in participation and readership on Wikipedia. External initiatives such as Mozilla's Digital Skills Observatory~\cite{mozillaskills}, a research project studying the impact of digital skills training on confidence and agency of low-income first-time smartphone users, provide useful insight in how to bridge this gap.
\subsubsection{Disabilities}
The \textit{disabilities} gap reflects how individual disabilities might affect one's ability to access the knowledge within Wikipedia. While individuals who are blind might be the most salient example, disabilities fall into many categories: cognitive, developmental, intellectual, mental, physical, or sensory disabilities. For some, disability can connote a cultural community as well. Access is closely tied to what options are available for interacting with Wikimedia projects (see Barriers below).
There is scant literature on the degree to which individuals who read Wikipedia have various disabilities beyond anecdotal evidence---e.g., an interview with Graham Pierce, an individual who is blind and a prolific editor~\cite{grahampierce}, tips for improving accessibility~\cite{signpost20130904}, a survey from 2008 that indicated that Wikipedia was a popular site for individuals who use screen readers~\cite{webaimscreenreadersurvey}. A number of projects focus on improving the accessibility of Wikimedia sites though, such as WikiProject Accessibility~\cite{wikiprojectaccess}, WikiProject Usability~\cite{wikiprojectusability}, WikiBlind User Group~\cite{wikiblind}, and Para-Wikimedians User Group~\cite{parawikimedians}.

\subsection{Barriers}\label{sec:readerbarriers}
We identified a few components which act as barriers preventing people from accessing free knowledge in Wikimedia projects. While this is not an comprehensive list, the following elements emerged during our literature review as potential causes and barriers for readers' knowledge gaps.

\subsubsection{Censorship}
Throughout the years, Wikipedia has been censored in various times at different locations around the world~\cite{clark2017analyzing}, and as of 2020, the access to the site is still banned in a few countries around the world~\cite{censorshipwiki}.
Even in countries where Wikipedia is not blocked, self-censorship plays a role in preventing people from accessing free knowledge: research has found that, as a result of the fear of surveillance, traffic to privacy-sensitive Wikipedia articles has been progressively declining~\cite{penney2016chilling}.
Due to country-wide blocks of Wikipedia and self-censorship, many readers therefore face difficulties accessing Wikipedia that generally require complicated technical solutions such as VPNs.

\subsubsection{Internet Connectivity and Cost}
Even though a given Wikipedia article is relatively small in size, slow internet connections can still make it very difficult to reach and browse Wikipedia. In surveys linking internet connectivity to readership~\cite{2010pewsurvey,2011phonesurvey,2016phonesurvey}, we see much lower readership among low-speed internet users. Evidence from early experiments after the introduction of a new data center in Singapore showed that lower latency correlates with higher longer-term reader engagement~\cite{phabT222078}. Internet cost can also be a huge barrier to internet access even when a country has good connectivity~\cite{2014globalsouthsurvey}. The data on usage after Wikipedia Zero program ended suggest that in places where data have indeed a high cost, it is a significant barrier to readership. Initiatives such as Kiwix~\cite{kiwix} and other projects from the Inuka team at the Wikimedia Foundation~\cite{newreadersresearch} aim at breaking this barrier by making Wikipedia available in offline settings.

\subsubsection{Device}
With predominantly mobile usage in countries that are only now coming online~\cite{2016phonesurvey,2014globalsouthsurvey}, making Wikipedia more accessible via mobile phones is a high priority. Efforts are already underway to support KaiOS\footnote{\url{https://www.kaiostech.com/}} (lightweight phone OS) and continued development of the mobile interface as well as official Android and iOS apps is important for welcoming this increasingly mobile population~\cite{kaios}.

\subsubsection{Interface Complexity}
One of the typical causes of interaction gaps such as \textit{Tech Skills} or \textit{Disabilities} is the complexity of the interface on Wikimedia sites.  
In general, providing a variety of ways to access content---e.g., text, images, video, audio---while reducing barriers to understanding---e.g., good color contrast, high readability---is a good approach to improving the accessibility for readers of all abilities~\cite{universaldesignlearning}.
The Design Research at the Wikimedia Foundation has developed a set of Wikimedia design guidelines~\cite{designwmf} that are based on the principle of ``designing for everyone''. This is aligned with the Wikimedia Strategy Recommendation of ``improving user experience''~\cite{uxstrategy} by making Wikimedia platforms more inclusive, allowing everyone to access free knowledge.
Initiatives aiming at breaking this barrier include, for example, Wikispeech~\cite{wikispeech} which provides text-to-speech for Wikipedia articles and VideoWiki~\cite{videowiki}, providing a tool for collaboratively editing videos from images and wikitext.

\subsubsection{Text Readability}
Content might be structured in such a way that it is inaccessible to readers with various cognitive, developmental, intellectual, mental, physical, or sensory disabilities---e.g., blindness, dyslexia. For English Wikipedia, the Accessibility component of the Manual of Style~\cite{manualofstyle} was written to address these issues and the Accessibility Dos and Don'ts~\cite{dosanddonts} provides a quick overview of some of the more prominent accessibility gaps in content. In particular, the following barriers are identified (to which we add how they might be better tracked):
\begin{itemize}
    \item Do use high contrast and color-blind friendly color schemes: \{\{Template:Overcolored\}\}, \{\{Template:Cleanup colors\}\}, and \{\{Template:Overcoloured\}\}, all of which add articles to [[Category:Wikipedia articles with colour accessibility problems]].
    \item Do provide alt text and a caption for most images: \{\{Template:Alternative text missing\}\} adds articles to [[Category:Unclassified articles missing image alternative text]], but this should also be detectable from wikitext dumps.
    \item Do provide a text description of any charts or diagrams: for tables, this should be detectable via wikitext dumps.
    \item Do nest section headings sequentially: pseudoheadings would be harder to detect but finding articles where section headers are out of order should just require going through parsed wikitext (or parsing the wikitext dumps with \textit{mwparserfromhell}\footnote{\url{https://mwparserfromhell.readthedocs.io/en/latest/}} or similar libraries).
    \item Do create correctly structured tables: it should be possible to detect lack of headers / scope for tables but other issues such as tables for layout or incorrectly structured tables would be much harder to detect.
    \item Do encase non-English words or phrases in \{\{lang\}\}: \{\{Template:Cleanup lang\}\} tracks instances where the \{\{lang\}\} template should be used and adds them to [[Category:Pages with non-English text lacking appropriate markup]]. There are also machine-learning models that do language identification but they are unlikely to work well for identifying the short stretches of other-language text for which these templates tend to be used.
\end{itemize}

%% file: tables/readership_table.tex
\begin{table}[]
\resizebox{\linewidth}{!}{
\renewcommand{\arraystretch}{1.2}
\begin{tabular}{p{3.2cm}|p{1.8cm}p{5.5cm}p{4.8cm}}
  \textit{\textbf{Facet}} &
  \textit{\textbf{Gap}} &  \textit{\textbf{Description}} &
  \textit{\textbf{Sources}}\\ \hline
\multirow{6}{3.4cm}{\textbf{~{ }\\Representation }\\~{ }
}
&&&\\
&
  \textit{Gender} &
  Difference between readers of different gender identities in how and how much they access the sites.  &
  literature~\cite{shaw2018pipeline,protonotarios2016similar,Hinnosaar2019newmedia}, surveys~\cite{2014globalsouthsurvey,2015donationsurvey,2016phonesurvey,2017strategysurvey,2019survey,2012banglasurvey,2014pakistansurvey,2016welshsurvey,2010pewsurvey}, strategy~\cite{uxstrategy,strategy}  
   \\
 &
  \textit{Age} &
  Difference between readers of different age in how and how much they access the sites.&
  literature~\cite{shaw2018pipeline,protonotarios2016similar,Hinnosaar2019newmedia},  surveys~\cite{2014globalsouthsurvey,2015donationsurvey,2016phonesurvey,2017strategysurvey,2019survey,2012banglasurvey,2014pakistansurvey,2016welshsurvey,2010pewsurvey}, strategy~\cite{uxstrategy,strategy,gendergapinitiatives}, community~\cite{simplewikiabout,wikibooksjunior,wikiprojectaccess} 
   \\
 &
  \textit{Geography} &
  Differences in readership between different areas of the world &
  literature~\cite{shaw2018pipeline},  surveys~\cite{2019survey,2016phonesurvey,2017strategysurvey}, community~\cite{wikiconnect,newreadersresearch}  
   \\
 &
  \textit{Language} &
  Differences in readership  depending on  readers' ability to read one or more languages &
  surveys~\cite{2019survey,2016welshsurvey,2014globalsouthsurvey,2011phonesurvey}, strategy~\cite{strategy,uxstrategy}, community~\cite{scribe,gapfinder}  
   \\
 &
  \textit{Socioeconomic Status} &
  Differences in readership depending on readers' education, income, wealth, or employment status &
  literature~\cite{shaw2018pipeline,Hinnosaar2019newmedia} surveys~\cite{2007pewsurvey,2019survey,2014pakistansurvey,2016welshsurvey,2014globalsouthsurvey,2008survey,2010pewsurvey,2012banglasurvey,2011phonesurvey,2011survey}, community~\cite{wikiversity,wikimediaeducation}   
\\
&
  \textit{Sexual Orientation} &
  Difference between readers of different sexual orientations in how and how much they access the sites. &
  literature~\cite{wexelbaum2015queering} 
  \\
&

  \textit{Cultural Background} &
  Differences in readership among people with different ethnic, political, and religious backgrounds &
  community~\cite{lgbtportal,npov} 
\\
&&&
   \\\hline
\multirow{5}{3.2cm}{\textbf{~{ }\\Interaction }\\~{ }
}
&&&\\
  &\textit{Motivation} &
  Differences in readership depending on the reason behind readers' visit to the site &
  literature~\cite{lemmerich2019world,2017motivationsurvey}, surveys~\cite{2019survey} 
   \\
  &\textit{Information Depth} &
  Differences in readership depending on the depth of information for which a reader is looking &
  literature~\cite{lemmerich2019world,singer2017we}, surveys~\cite{2019survey,2019trustsurvey} 
 \\
 &
  \textit{Familiarity} &
  Differences in readership depending on one's prior familiarity with a topic &
  literature~\cite{lemmerich2019world,2017motivationsurvey}, surveys~\cite{2019survey}, community~\cite{breadthdepth}  
  \\
 &
  \textit{Tech Skills} &
  Differences in readership depending on readers' general internet skill &
  literature~\cite{shaw2018pipeline}, strategy~\cite{uxstrategy}, other~\cite{mozillaskills}  
   \\
 &
  \textit{Disabilities} &
  Disparities in ability to access the knowledge within Wikipedia depedning on individual disabilities &
  literature~\cite{webaimscreenreadersurvey}, community~\cite{grahampierce,signpost20130904,wikiprojectaccess,wikiprojectusability,wikiblind,parawikimedians,wikispeech,videowiki} 
  \\
&&&
  \\\hline
\end{tabular}}
\end{table}

%% file: sections/4_b_Contributors_Taxonomy.tex
\input{tables/contributor_table}

\section{Contributors}
\label{sec:contributors}
The contributor dimension of knowledge gaps covers all gaps related to categories of people contributing to Wikimedia sites. We define contributors as all individuals who \textit{edit} or otherwise \textit{maintain} Wikimedia content. For the purpose of this taxonomy, this definition does not include technical contributors---i.e. the individuals who build the MediaWiki software on which Wikimedia sites run---though the software and choices made in its design certainly are highly impactful on what types of contributors feel supported and what content is created~\cite{ford2017anyone}. It also does not specifically address bot accounts, though the different roles these accounts take on can have a large impact on the community and content~\cite{halfaker2012bots,zheng2019roles,johnson2016not}.
Contributors gaps are organized into two main facets: \textit{representation} gaps---i.e. who is included in the contributor population---and \textit{interaction} gaps---i.e. are individuals supported regardless of their context or ways in which they would like to interact with Wikimedia. We also highlight different initiatives aiming at closing each gap and some common barriers to greater equity in this space. 
Similar to the Readership dimension, the gaps in the Contributors facet are inspired by the 2020 Movement Strategy, which invites Wikimedia communities to be safe and inclusive~\cite{inclusivestrategy} and projects to be designed so that everyone is welcome to contribute~\cite{uxstrategy}.
\subsection{Representation Gaps}
Similar to their Readers counterpart, contributors' representation gaps---i.e. sociodemographic gaps such as gender, age, and education---have been widely discussed by different sources.
\subsubsection{Gender}
The \textit{gender} gap is the difference between individuals of different gender identities in how likely they are to contribute to Wikimedia sites. The gender gap has been measured through many surveys organized by the Wikimedia Foundation~\cite{2011survey,2011phonesurvey,2011aprileditorsurvey,2011novembereditorsurvey,2012editorsurvey,2014globalsouthsurvey,2017strategysurvey,2018communityinsights,2017communityinsights,2019editorgendersurvey,2019yougov,2020communityinsights}, different chapters~\cite{2012banglasurvey,2014pakistansurvey,2016wmde,2016ireland,2018ukraine,2015wikimedianl,2013dutchsurvey,merz_wess2011}, independent organizations~\cite{2008survey} and research communities~\cite{shaw2018pipeline,protonotarios2016similar,Hinnosaar2019newmedia,hanna2014motivate} by asking respondents for their gender identity, though generally limiting the possible answers to just a binary man/woman choice and do not specify between cis and trans identities. Conclusions about gender gaps might differ depending on the definition of contributors. For example, recent surveys in this space~\cite{2019editorgendersurvey} stratified editors by how many edits were associated with their account. For all three languages surveyed---Arabic, English, and Norwegian---the gender gap is less extreme for editors with fewer edits.
While researchers are still working on investigating the causes behind some gender identities being under-represented in Wikimedia's contributor population, different community initiatives are focusing on bridging this gap~\cite{gendergapinitiatives}. 
\subsubsection{Age}
The \textit{age} gap is the difference between individuals of different age in how likely they are to contribute to Wikimedia sites.
Contributors' age data has been collected by surveys and academic literature~\cite{2008survey,2011survey,2011phonesurvey,2011aprileditorsurvey,2011novembereditorsurvey,2012editorsurvey,2012banglasurvey,Hinnosaar2019newmedia,2014pakistansurvey,2014globalsouthsurvey,protonotarios2016similar,shaw2018pipeline,2016wmde,2016ireland,2017strategysurvey,2017communityinsights,2018communityinsights,2018ukraine,2015wikimedianl,2020communityinsights,2013dutchsurvey,merz_wess2011,hanna2014motivate}. These studies found that the median age of contributors varies substantially, but the median age tracks slightly lower than country-by-country median age.\footnote{\url{https://en.wikipedia.org/wiki/List_of_countries_by_median_age}} This is best captured by the 2018 Community Insights survey~\cite{2018communityinsights}. Median age ranges from around 40 in Western Europe~\cite{2015wikimedianl,2016wmde,2017communityinsights,2018communityinsights,2016ireland} to the 20s in Eastern Europe~\cite{2018ukraine,protonotarios2016similar} and ``Global South''~\cite{2014globalsouthsurvey}. Globally, the median age has increased overtime from the low 30s~\cite{2011novembereditorsurvey,2012editorsurvey} to high 30s~\cite{2017communityinsights,2018communityinsights}. Throughout the years, several initiatives have been developed to bridge this gap and involve senior citizens as contributors for Wikimedia projects~\cite{seniorsoutreach}.

\subsubsection{Geography}
The \textit{geography} gap is the difference between where an individual lives---both which country and where in the country---and how likely they are to contribute to Wikimedia sites. Understanding editors' locales is quite important to understanding gaps as it is both relatively easy to infer with high accuracy and also highly indicative of their cultural context and availability of services. In particular, there is an abundance of data about which countries editors come from because country can be inferred with high accuracy based on an editor's IP address.\footnote{Which country an editor's IP address is from is different than where they live or are from, but generally it is assumed that there is high overlap.} While that information is not public (unless the individual was not signed-in when they edited), the geoeditors dataset~\cite{geoeditors} provides monthly snapshots of about how many editors from each country contributed to each project. Analyses of this and related data have examined the relationship between what regions the content is about on Wikipedia and where the editors who wrote that content are from~\cite{sen2015barriers,yasseri2012circadian}. Various surveys~\cite{2008survey,2011aprileditorsurvey,2011novembereditorsurvey,2012editorsurvey,2017communityinsights,2018communityinsights,2020communityinsights,2012banglasurvey,2015wikimedianl} have also been able to explore the distribution of editors for various language editions without requiring access to the more sensitive IP information.
\subsubsection{Language}
The \textit{language} gap is the difference between an individual's fluency in a language and how likely they are to contribute to Wikimedia sites.
Surveys have been conducted to estimate contributors' literacy or language skills~\cite{2011phonesurvey,2014globalsouthsurvey,2016ireland,2018ukraine,2015wikimedianl,2020communityinsights} and the Babel system~\cite{babelsystem} is widespread on user talk pages and offers an alternative to understanding the fluency of contributors. Though it may feel intuitive that fluency would be required to contribute, lowering the barrier to contribution by lower-fluency individuals can be important for effective patrolling in small wikis~\cite{patrollingwikimedia}, increase the diversity of contributors, and allow for the cross-pollination of content that might otherwise remain locked up in other languages~\cite{hale2014multilinguals}. Many editors are multilingual and contribute to Wikipedia in a variety of languages, with small wikis heavily depending on multilingual editors and English the most common second-language outside of one's native language~\cite{2015wikimedianl,hale2014multilinguals,2014globalsouthsurvey}. While reducing language barriers is important, it also brings risks of larger communities overshadowing the contributions of more local contributors as happened recently with Scots Wikipedia~\cite{scotswikirewrite}. Tools like Scribe~\cite{scribe} have sought to address the language gap by making it easier to contribute in one's own language even when there are not easier approaches to writing articles like Content Translation~\cite{contenttranslation} available.
\subsubsection{Socioeconomic Status}
The \textit{socioeconomic status} gap is the difference between individuals of different education, income, wealth, or employment and how likely they are to contribute to Wikimedia sites. Research on contributor income~\cite{2011phonesurvey,shaw2018pipeline} and employment~\cite{2011survey,2011phonesurvey,2011aprileditorsurvey,2011novembereditorsurvey,2012editorsurvey,2012banglasurvey,Hinnosaar2019newmedia,2014pakistansurvey,2014globalsouthsurvey,2016ireland,2018ukraine} does not show consistent patterns in contribution by income~\cite{shaw2018pipeline} but higher contribution rates among individuals who are employed~\cite{2014globalsouthsurvey,shaw2018pipeline}. This is presumably convoluted by the high number of student contributors to Wikipedia~\cite{2012editorsurvey}. Regarding education, several surveys~\cite{2008survey,2011survey,2011phonesurvey,2011aprileditorsurvey,2011novembereditorsurvey,2012editorsurvey,2012banglasurvey,Hinnosaar2019newmedia,2014pakistansurvey,2014globalsouthsurvey,shaw2018pipeline,2016ireland,2018ukraine,2018communityinsights,2015wikimedianl,2020communityinsights,merz_wess2011,hanna2014motivate} consistently demonstrate that individuals with higher levels of education are more likely to be contributors to Wikipedia. Despite this skew, many academics, students, librarians, and other scholars do not contribute to Wikipedia. Organizations like Wiki Education~\cite{wikiedu} and the Wikipedia Library~\cite{wikipedialibrary} work to change this.

\subsubsection{Sexual Orientation}
The \textit{sexual orientation} gap is the difference between contributors who hold various sexual orientations and how they contribute to the site. Cultural communities around sexual orientation often are tied to gender identity as well as with LGBTQ+ communities.\footnote{\url{https://meta.wikimedia.org/wiki/Wikimedia_LGBT\%2B/Portal}} There has been less empirical research that has specifically focused on sexual orientation on-wiki within the contributor community, but plenty of evidence of barriers facing this community. Surveys as early as 2011~\cite{2011aprileditorsurvey} showed that editors were being harassed about their sexual orientation and the Wikimedia LGBT+ user group~\cite{lgbtportal} was beginning to organize as early as 2006~\cite{wexelbaum2015queering}. There are WikiProjects focused on LGBT+ issues on at least 29 wikis\footnote{\url{https://www.wikidata.org/wiki/Q15092984}} and the Wikimedia Foundation released a statement in December 2020 detailing issues around safety faced by these communities~\cite{lgbtsupport}. Notably, given the oppression around sexual orientation in many countries around the world, extreme care should be taken before including questions about sexual orientation in surveys to contributors.
\subsubsection{Cultural Background}
The \textit{cultural background} gap includes differences in contributorship among people with different ethnicity, race, political beliefs, or religious beliefs. These are gaps that tend to be very contextual and specific to regions. Ethnicity and race, or other proxies for caste, are very contextual as to what it means about an individual's status and access to resources. Studies in the United States have shown gaps in contribution along racial lines~\cite{shaw2018pipeline} and presumably these same trends are found in other countries where groups of people who are disenfranchised due to their race/ethnicity are also less likely to contribute to Wikipedia.

No comprehensive surveys were found that included measures of individual beliefs related to politics, religion, or culture such as those asked by the European Values Survey~\cite{evs2017}. Research~\cite{shi2019wisdom,keegan2019dynamics} has attempted to assign political leanings to editors, however, based upon their contribution history, showing interest in how individual beliefs of contributors affect the process of collaboration, the degree to which they contribute, and what articles they edit. One of the main recommendations within the 2030 movement strategy encourage communities to bridge these gaps by making the community more inclusive and safe~\cite{inclusivestrategy}. One of the main mechanisms the community uses to ensure respect for the diversity of identities and beliefs represented is the Friendly Space Policy, which is in force at every community event~\cite{friendlyspace}. Work is ongoing regarding a Universal Code of Conduct to establish basic expectations of acceptable behavior and enforcement for the entire movement as well~\cite{universalcoc}.

\subsection{Interaction Gaps}
Individuals contribute to Wikipedia for many different reasons and in many different ways. Our tools need to support this diversity of motivations, skills and types of work or we risk exacerbating gaps by only supporting certain types of contributors. Of note, while research that seeks to reduce barriers to contributing (e.g., socialization via Wikipedia Teahouse~\cite{morgan2018evaluating}, personalized edit recommendations from SuggestBot~\cite{cosley2007suggestbot} or GapFinder~\cite{wulczyn2016growing}) has been demonstrated to increase contributions, research that has focused on playing directly to editors motivations is a more cautionary tale and can instead supersede existing motivations (see Gamified Onboarding~\cite{narayan2017wikipedia}), or be perceived as manipulative~\cite{rolespecificrewards}.

\subsubsection{Motivation}
Motivation gaps reflect different levels of contributions depending on the reason behind an individual's desire to contribute to the site. Contributor motivations include both the intrinsic (e.g., support of free knowledge, for fun, curiosity, personal satisfaction) and extrinsic (e.g., fixing errors, promoting a topic, professional or school-related reasons, learning a new skill, money)~\cite{hanna2014motivate}. This wide range of motivations affect contributors' levels of activity~\cite{algan2013cooperation,balestra2016motivational} and roles that they take on~\cite{arazy2017and}. Contributors' initial motivations\footnote{\url{https://upload.wikimedia.org/wikipedia/commons/8/8a/Global_South_User_Survey_2014_-_Full_Analysis_Report.pdf?page=235}; \url{https://upload.wikimedia.org/wikipedia/commons/7/76/Editor_Survey_Report_-_April_2011.pdf?page=12}} can also differ greatly from the reasons they stay.\footnote{\url{https://upload.wikimedia.org/wikipedia/commons/8/81/Editor_Survey_2012_-_Wikipedia_editing_experience.pdf?page=19}; \url{https://upload.wikimedia.org/wikipedia/commons/7/76/Editor_Survey_Report_-_April_2011.pdf?page=14}} See ``why do people edit?''~\cite{Rader2020whydopeopleedit}, Balestra et al.~\cite{balestra2017fun}, various surveys~\cite{2008survey,2011aprileditorsurvey,2014globalsouthsurvey,2015wikimedianl}, or the detailed results\footnote{\url{https://upload.wikimedia.org/wikipedia/commons/a/a1/Wikipedia_Editor_Survey_2012_-_motivation_analysis.pdf}} from the 2012 Editor Survey~\cite{2012editorsurvey} for additional details.
\subsubsection{Role}
Roles reflect different levels of contributions depending on what type of work an individual does to support the wikis. Editors might be dedicated to one or more projects, and have different levels of experience and lifetime on their projects. There are explicit roles---i.e. user access levels~\cite{useraccesslevels}---on Wikipedia through which users ascend~\cite{arazy2015functional}. However, within these access levels, individuals can do very different types of work~\cite{contributiontaxonomy}. Yang et al.~\cite{yang2016did} identified seven editor roles that users of all access levels undertake. These can be expanded to more specific types of work but offer a high-level view and can be modeled from edit activity. 
\begin{itemize}
    \item Social Networker: communicate on user pages and communication namespaces
    \item Fact Checker: removal / verification of content
    \item Substantive Expert: content producers---adding substantive content to articles
    \item Copy Editor: grammar, paraphrase, relocation
    \item Wiki Gnomes: clean up content and wikitext markup issues
    \item Vandal Fighter: reverting vandalism, warning editors, and other patrolling work
    \item Fact Updater: updating templated content or Wikidata
    \item Wikipedian: working behind the scenes (non-article mainspace) to keep things organized etc.
\end{itemize}
Additionally one might also include the following roles that do not necessarily have public traces on Wikimedia but are highly important roles that contributors also take: organizers, public advocates, technical contributors, affiliates, board members.  
Surveys generally ask contributors what types of work they do on-wiki~\cite{2011aprileditorsurvey,2011novembereditorsurvey,2012editorsurvey,2014globalsouthsurvey,2017communityinsights}, but these tasks can be matched to the above roles. Initiatives such as the Wikipedia Teahouse~\cite{morgan2018evaluating} or the Newcomer Page~\cite{newcomerhomepage} help newer Wikipedians learn how to familiarize themselves with the platform and progressively become more expert editors.
\subsubsection{Tech Skills}
The \textit{tech skills} gap reflects different levels of contributions depending on an individual's general internet skills. While extensive effort has been put into simplifying the process of contributing to Wikipedia, editing articles, uploading images, and identifying sources can still pose substantial technical challenges for individuals.
Academic research~\cite{shaw2018pipeline} has found that high internet skills are associated with an increase in awareness that Wikipedia can be edited and having edited Wikipedia. Edit-a-thons\footnote{\url{https://en.wikipedia.org/wiki/Edit-a-thon}} in particular can help to bridge this gap and the 2030 strategy calls for further investment in skills and training~\cite{skillsstrategy}.
\subsubsection{Disabilities}
The \textit{disabilities} gap reflects how individual disabilities might affect one's ability to access and contribute to the knowledge within Wikipedia. While individuals who are blind might be the most salient example, disabilities fall into many categories: cognitive, developmental, intellectual, mental, physical, or sensory disabilities. There is scant evidence beyond the anecdotal of the degree to which individuals with various disabilities contribute to Wikipedia (see Userboxes in WikiProject Accessibility~\cite{wikiprojectaccess}, an interview with Graham Pierce~\cite{grahampierce}, and evaluation from 2008 of editing with a screenreader~\cite{buzzi2008making}). Various groups have also been established for individuals with disabilities such as the WikiBlind User Group~\cite{wikiblind} and Para-Wikimedians User Group~\cite{parawikimedians}.

\subsection{Barriers}\label{sec:contribbarriers}
Barriers and causes of Contributors' knowledge gaps match most of the ones we identified for readers in Sec.~\ref{sec:readerbarriers}. 
\subsubsection{Censorship}
Similar to the considerations around censorship and readers in Sec.~\ref{sec:readerbarriers}, both direct and self-imposed censorship can be important barriers preventing people to contribute to the sites. In a study around Wikipedia censorship in China, Zhang and Zhu~\cite{zhang2011group} found that blocks to Chinese Wikipedia in Mainland China in October 2005, beyond preventing contributions from individuals located in Mainland China, also led to a 42.8\% drop in contributions from non-blocked individuals.

\subsubsection{Internet Connectivity}
As seen in the case of readership, the quality of internet access can also affect one's ability to contribute. This covers both internet speed and internet cost. While Wikipedia content is relatively lightweight, cost and speed have been identified as a major barriers to contributors~\cite{2014globalsouthsurvey}. Graham et al.~\cite{graham2014uneven} found that removing broadband access barriers is a necessary but insufficient condition for content generation. Initiatives like Growing Local Language Content on Wikipedia~\cite{glow} work on addressing this gap by subsidizing internet costs for contributors.

\subsubsection{Device}
Editing Wikimedia projects classically requires a good laptop or desktop computer in order to easily use the available editing interfaces while potentially also gathering research in other windows. There has been a shift to mobile that has been accompanied with huge improvements in mobile editing interfaces (e.g., Suggested Edits~\cite{suggestededits}). Additionally, efforts have been made to provide laptops (e.g., GLOW~\cite{glowfaq}) to new editors in order to reduce this technological burden. Various surveys~\cite{2011survey,2011phonesurvey,2011aprileditorsurvey,2011novembereditorsurvey,2012editorsurvey,2014globalsouthsurvey} have asked contributors about their devices.

\subsubsection{Interface Complexity}
Interaction gaps such as Technical Skills and Disabilities are largely influenced by the complexity of the interface. To reduce the technical entry barriers for Wikipedia editing, in 2013 the Editing team at the Wikimedia Foundation introduced the VisualEditor~\cite{visualeditor}, a rich-text editor now available on almost all Wikipedias and Wikivoyages. While the visual editor has been found to have no effect on new editor retention or productivity, it helped creating more reliable (less reverted) edits \cite{visualeditoreffects}. Another project to reduce interface complexity is the ``Talk Pages Project'', which aims at adding functions such as direct responses to Wikipedia discussion pages.
Beyond Wikipedia, initiatives such as Structured Data on Commons~\cite{sdc} also aimed at removing technical barriers to contribution to Wikimedia Commons, by introducing an easy interface for image description and categorization.

\subsubsection{Privacy}
Another important barrier for contribution is the importance that different people give to privacy. 
Many contributors have strong privacy concerns around their contributions to Wikipedia. Usernames allowed much of one's identity to be protected, and Wikimedia's privacy policy ensures that contributors should not disclose ``nonpublic personal information to participate in the free knowledge movement''~\cite{privacypolicy}. 
However, some individuals have credible reasons to contribute while obscuring their IP address, even from private server logs. These individuals, while possibly making equally high-quality contributions~\cite{tran2020anonymity}, face the challenge that Tor and many of these proxy IP addresses~\cite{ipblockexemption,openproxies} are blocked due to vandals (or perceived potential for vandalism) using them to circumvent IP blocks.

\subsubsection{Community Health}
A well-documented barrier to retaining a diverse community of contributors is the lack of inclusivity and sometimes outright harassment (lack of safety) that some experience in the editor community. This has been identified as a major challenge for newcomer retention~\cite{halfaker2013rise,morgan2018evaluating} and been raised consistently as a challenge in community surveys~\cite{2017communityinsights,2018communityinsights,2015harassmentsurvey,2011aprileditorsurvey,2011novembereditorsurvey,2012editorsurvey,2020communityinsights,communityhealth}. This harassment can be disproportionately targeted at already-marginalized communities--e.g., women~\cite{menking2015heart,2020communityinsights,ford2017anyone,genderequity2018}. Initiatives like the Community Health Initiative~\cite{communityhealthinitiative} and Universal Code of Conduct~\cite{universalcoc} seek to address some of these challenges by establishing clearer norms, processes, and better tools for identifying and dealing with harassment.

%% file: tables/contributor_table.tex
\begin{table}[]
\resizebox{\linewidth}{!}{
\renewcommand{\arraystretch}{1.2}
\begin{tabular}{p{3.2cm}|p{1.8cm}p{5.5cm}p{4.8cm}}
  \textit{\textbf{Facet}} &
  \textit{\textbf{Gap}} &  \textit{\textbf{Description}} &
  \textit{\textbf{Sources}}\\ \hline
\multirow{6}{3.4cm}{\textbf{~{ }\\Representation }\\~{ }
}
&&&\\
&
  \textit{Gender} &
  Differences between contributors of different gender identities in how and how much they contribute to the sites.  &
  literature~\cite{shaw2018pipeline,protonotarios2016similar,Hinnosaar2019newmedia}, surveys~\cite{2011survey,2011phonesurvey,2011aprileditorsurvey,2011novembereditorsurvey,2012editorsurvey,2014globalsouthsurvey,2017strategysurvey,2018communityinsights,2017communityinsights,2019editorgendersurvey,2019yougov,2008survey,2020communityinsights}, strategy~\cite{uxstrategy,strategy}, community~\cite{2012banglasurvey,2014pakistansurvey,2016wmde,2016ireland,2018ukraine,2015wikimedianl}
   \\
 &
  \textit{Age} &
  Differences between contributors of different ages in how and how much they contribute to the sites.&
  literature~\cite{shaw2018pipeline,protonotarios2016similar,Hinnosaar2019newmedia},  surveys~\cite{2008survey,2011survey,2011phonesurvey,2011aprileditorsurvey,2011novembereditorsurvey,2012editorsurvey,2014globalsouthsurvey,2017strategysurvey,2017communityinsights,2018communityinsights,2020communityinsights}, strategy~\cite{uxstrategy,strategy,gendergapinitiatives}, community~\cite{2012banglasurvey,2014pakistansurvey,2016wmde,2016ireland,2018ukraine,2015wikimedianl,seniorsoutreach}
   \\
 &
  \textit{Geography} &
  Differences between contributors of different geographies in how and how much they contribute to the sites. &
  literature~\cite{shaw2018pipeline,johnson2016not},  surveys~\cite{2017strategysurvey,2020communityinsights}, community~\cite{2016ireland,wmspain_ap2019,wikilovesvillages}
   \\
 &
  \textit{Language} &
  Differences between contributors of different reading abilities in a language in how and how much they contribute to the sites. &
  surveys~\cite{2011phonesurvey,2014globalsouthsurvey,2020communityinsights}, strategy~\cite{strategy,uxstrategy}, community~\cite{2016ireland,2018ukraine,2015wikimedianl,babelsystem} 
   \\
 &
  \textit{Socioeconomic Status} &
  Differences between contributors of different education, income, wealth or employment status in how and how much they contribute to the sites. &
  literature~\cite{Hinnosaar2019newmedia,shaw2018pipeline}, surveys~\cite{2008survey,2011survey,2011phonesurvey,2011aprileditorsurvey,2011novembereditorsurvey,2012editorsurvey,2014globalsouthsurvey,2018communityinsights,2020communityinsights}, community~\cite{2012banglasurvey,2014pakistansurvey,2016ireland,2018ukraine,2015wikimedianl,wikiedu}
  \\
  &
  \textit{Sexual Orientation} &
  Differences between contributors of different sexual orientations in how and how much they contribute to the sites. &
  literature~\cite{genderequity2018,wexelbaum2015queering},
  surveys~\cite{2011aprileditorsurvey},
  strategy~\cite{lgbtsupport},
  community~\cite{lgbtportal}
  \\
  &
  \textit{Cultural Background} &
  Differences between contributors of different political, religious, or ethnic background in how and how much they contribute to the sites. &
  literature~\cite{shaw2018pipeline,shi2019wisdom,keegan2019dynamics} surveys~\cite{evs2017}, community~\cite{whoseknowledge,afrocrowd,friendlyspace,inclusivestrategy}
  \\
&&&
   \\\hline
\multirow{4}{3.2cm}{\textbf{~{ }\\Interaction }\\~{ }
}
&&&\\
  &\textit{Motivation} &
  Differences in contribution depending on one's reason for contributing to the site. &
  literature~\cite{algan2013cooperation,balestra2016motivational,arazy2017and,balestra2017fun,Rader2020whydopeopleedit}, surveys~\cite{2014globalsouthsurvey,2011aprileditorsurvey,2012editorsurvey,2008survey,2015wikimedianl}
   \\
 &
  \textit{Role} &
  Differences in contribution depending on the type of editing that one chooses to do. &
  literature~\cite{arazy2015functional,yang2016did}, surveys~\cite{2011aprileditorsurvey,2011novembereditorsurvey,2012editorsurvey,2014globalsouthsurvey,2017communityinsights}, community~\cite{useraccesslevels,morgan2018evaluating,newcomerhomepage}
  \\
 &
  \textit{Tech Skills} &
  Disparities in ability to contribute to the knowledge within Wikipedia depending on one's general internet skills &
  literature~\cite{shaw2018pipeline}, strategy~\cite{skillsstrategy,uxstrategy}  
   \\
 &
  \textit{Disabilities} &
  Disparities in ability to contribute to the knowledge within Wikipedia depending on individual disabilities &
  literature~\cite{buzzi2008making}, community~\cite{wikiblind,parawikimedians,wikiprojectaccess,grahampierce}
  \\
&&&
  \\\hline
\end{tabular}}
\end{table}

%% file: sections/4_c_Content_Taxonomy.tex
\input{tables/content_table}

\section{Content}
\label{sec:content}
Wikipedia is incomplete by design. The opportunity to share new information with the world is a major motivating factor among both new and established Wikipedia contributors. However, when important information about a topic is absent, incomplete, biased, or otherwise inaccessible to readers, these content gaps can undermine Wikipedia's ability to serve the needs of its global audience. 

The goal of this Section is to characterize \textit{gaps in content coverage}. In the widest sense, ``content'' is information about a topic, i.e. a piece of knowledge that could be the focus of one or more Wikipedia articles. ``Coverage'' refers to how well Wikimedia project content addresses a particular topic. In turn, a content gap refers to differences in coverage of one or more topics.

The gaps in this dimensions are largely inspired by the ``Impactful Topics'' Movement Strategy recommendations, which suggests that Wikimedia communities should ``develop and increase access to content that has historically been left out by structures of power and privilege''~\cite{topictrategy}. Furthermore, the facets have been inspired by community discussions on systemic biases~\cite{systemicbias} as well as reviews on Wikipedia's content~\cite{Mesgari2015sum} and its biases~\cite{Hube2017bias}.
\subsection{Representation Gaps}
The content representation facet includes all gaps related to topic coverage in Wikimedia projects. 
Developing a comprehensive, hierarchically-structured, and canonical representation of Wikipedia content is a non-trivial if not impossible task. Wikipedia's own category structure for content\footnote{\url{https://en.wikipedia.org/wiki/Wikipedia:Categorization}} is not hierarchical, and the language and culture-specific nature of topic relevance make the creation of an exhaustive list of topic gaps infeasible. Therefore, our aim here is to provide a list of example topic gaps that have received widespread attention in public media, research studies, and include topics highly curated by community projects and discussed across different parts of the movement, such as gender, geography, and culture gaps. Gaps in this facet are inspired by both the Wikimedia Foundation's Medium-term plan~\cite{mtp}, which supports diverse content creation ``by actively foster[ing] the inclusion of underrepresented and marginalized knowledge, and ensur[ing] content is locally relevant to communities'', as well as the the movement strategy directions~\cite{topictrategy} which encourages ``improving coverage of collectively-identified priority topics that impact our world and improve people's lives''. Community initiatives such as the Wikiproject Countering Systemic Bias~\cite{counteringsystemicbias}, the Wikipedia Diversity Observatory~\cite{miquel2019wikipedia}, or Whose Knowledge?~\cite{whoseknowledge} aim to address imbalances in the coverage of subjects and topics.
\subsubsection{Gender}
The \textit{gender} gaps refer to the differences in content coverage  depending on the gender identity of subjects. 
The gender gap---the fact that the majority of Wikimedia content about people focuses on male subjects---is one of the most-studied examples of content gaps and well-documented in the research literature ~\cite{Reagle2011britannica,Bamman2014unsupervised,Eom2015interactions,graells2015first,Wagner2015mans,Wagner2016glass,Young2016ceo,Halfaker2017keilana,Menking2017who,Zagovora2017professions, konieczny2018gender,Adams2019who,Hinnosaar2019newmedia,Schellekens2019recognition} and interactive tools such as the Wikidata Human Gender Indicators~\cite{whgi}, Denelezh~\cite{denelezh}, or the Wikidata Concepts Monitor~\cite{wdcm_bias}. A number of community initiatives such as the Gender Gap Portal~\cite{gendergapportal}, the Wikiproject Women~\cite{wikiprojectwomen}, as well as organizations such as Whose Knowledge?~\cite{visiblewikiwomen}, are focusing their efforts and working hard to address the gender content gap across Wikimedia projects.
 
\subsubsection{Age (Recency)}
The \textit{age (recency)} gap reflects the difference in coverage depending on the point in time a given topic, fact or event has taken place. That Wikipedia suffers from a ``recency'' bias is a well-known fact: researchers have found that, consistently across Wikipedia languages, content is skewed towards more recent events~\cite{samoilenko2017analysing}. This bias seems to match content demand. Academic research has shown that recent events attracts more views on Wikipedia~\cite{ciglan2010wikipop}, especially as a result of wide media coverage~\cite{tizzoni2020impact}. Recent work has also found that readers interact more often with content within the page which refers to recent events~\cite{piccardi2020quantifying}. To improve the availability of content across different points in time, and counter the systemic recency bias, some communities such as Wikiproject Historical Information~\cite{wikiprojecthistorical} focus on identifying and fixing articles that lack historical information.

\subsubsection{Geography}
The geographic gap captures coverage differences in topics related to geographic regions or population distribution. Looking at geo-tagged information on Wikipedia content, research has shown that the
geographic coverage differs substantially across language editions~\cite{Hecht2009selffocus,Beytia2020positioning} and that geographic coverage is extremely uneven and clustered with a strong bias towards content related to the United States and Western Europe~\cite{Graham2011geographies,Graham2014geo}. Within countries, there also tends to be a strong bias towards content in urban areas as opposed to rural areas~\cite{johnson2016not}. To address this gap, several initiatives across Wikimedia projects aim at increasing content coverage of under-represented areas such as the Africa Portal in Wikipedia~\cite{africaportal}, Wiki Loves Africa Contest in Wikimedia Commons~\cite{wikilovesafrica}, and Wiki Takes campaign to document rural Spain~\cite{wikitakes_rural}.
\subsubsection{Language}
The \textit{language} gap refers to the difference in content coverage across different languages. While each Wikipedia language edition is a stand-alone project, with different size and coverage of relevant topics~\cite{wulczyn2016growing,bao2012omnipedia}, other projects such as Wikidata and Wikimedia commons are multilingual by design. However, while Wikimedia Commons is used across many languages~\cite{he2018the_tower_of_babel}, its captions and descriptions area available mainly in English.  Wikidata's labels are also non-uniformly distributed across languages, with only 11 languages holding almost
50\% of all language knowledge in Wikidata, English being one of the most prominent ones~\cite{kaffee2017glimpse}. Projects such as Structured Data on Commons~\cite{sdc} and Suggested Edits~\cite{suggestededits} aimed at rehauling the projects' interface to make the translation efforts on Commons easier and more effective.
\subsubsection{Socioeconomic Status}
The \textit{socioeconomic status} gap refers to different levels of content coverage depending on the socioeconomic status of a location or a person. That content in Wikipedia is skewed towards western countries with higher Human Development Index is a well-established fact~\cite{wulczyn2016growing,rask2008reach}. Several projects such as the Cultural Diversity Observatory~\cite{miquel2019wikipedia} and Wikiproject Countering Systemic Bias~\cite{counteringsystemicbias} are building tools and empowering communities to foster the inclusion of content from marginalized regions into Wikimedia projects.
\subsubsection{Sexual Orientation}
The \textit{sexual orientation} gap is the difference in content relevant to various sexual orientations. Sexual orientation is often discussed in relation to gender identity as with LGBTQ+ communities.\footnote{\url{https://meta.wikimedia.org/wiki/Wikimedia_LGBT\%2B/Portal}} There has been less empirical research that has specifically focused on content and sexual orientation, but the portrayal of LGBTQ+ individuals has been shown to vary across different language editions~\cite{park2020multilingual}. Wikidata WikiProject LGBT~\cite{wikidatalgbt} has been working on expanding well-sourced statements about sexual orientation in Wikidata, which will greatly help content analyses around the representation of sexual identities.

\subsubsection{Important Topics}
The \textit{important topics} gap captures coverage differences among content deemed to be especially impactful to readers and important to be of high quality. The abundance of topics varies substantially across different Wikipedia language editions~\cite{Kittur2009whatsinwikipedia,Lewoniewski2019multilingual}. Certain topics may be deemed to take on special importance such as the coverage of medical knowledge on Wikipedia~\cite{Shafee2017evolution} since ``[its] health content is the most frequently visited resource for health information on the internet''~\cite{Smith2020health} (see also the WikiProject Medicine in English Wikipedia~\cite{wikiprojectmedicine}). While the Movement Strategy indicates that the community is still missing tools to identify which topics are most impactful in the world, initiatives such as WikiProject Vital Articles~\cite{vitalarticles} or All Human Knowledge~\cite{allhumanknowledge} help in addressing this gap by gathering lists of important and impactful topics that should be present in all Wikipedia editions.
\subsubsection{Cultural Background}
The \textit{cultural background} gap captures coverage differences related to the history, heritage, and characteristics of different cultural groups. Cultural identity has been shown to be a crucial part for the motivation of editors to contribute~\cite{MiquelRibe2016}, and the project Wikipedia Diversity Observatory~\cite{wcdo} aims to define and quantify the extent of articles which can be considered cultural context content~\cite{MiquelRibe2018}. Initiatives such as Wikiproject French Caribbean Culture~\cite{frenchcarribeanculture} or AfroCrowd~\cite{afrocrowd} aim to address some of these cultural gaps.

\subsection{Interaction Gaps}
Content in Wikimedia projects is multimodal by nature, and can take very different forms, such as images, text, structured data, etc., and the form of presentation clearly affects how accessible content is to different audiences~\cite{strategy}. The gaps in this facet cover various aspects of the content available in Wikimedia spaces that mediate how people access free knowledge and can help to break down barriers to access.
\subsubsection{Multimedia}
The \textit{multimedia} gap reflects differences in coverage with respect to the type of media used to share the content. Acknowledging the potential of different forms of media beyond text (image, audio, video, geospatial, etc.) to convey content to different audiences, the Movement strategy~\cite{innovationstrategy} recommends building the necessary technology to make free knowledge content accessible in various formats and support more diverse modes of consumption and contribution to Wikimedia projects. For example, with the bulk of a typical article made up of text, the use of images is encouraged in order to increase readers' understanding of the subject matter (see, for example, English Wikipedia's policy on image use~\cite{imageusepolicy}). 
However, on average, half of Wikipedia articles are missing images, and around 95\% of Wikidata items do not have a value for the image-property ($P18$)\footnote{\url{https://www.wikidata.org/wiki/Property:P18}}~\cite{imagewikipedia}. 
This is exacerbated by the fact that with few exceptions (e.g.,~\cite{Viegas2007,He2018babel,Porter2020visual}), the role of visual and multimedia aspects in Wikipedia and other projects has largely been ignored by researchers. 
To help in bridging this gap, several recent research initiatives~\cite{imageclassifiers,imagerole,imagewikidata,imagewikipedia} aim at designing smart tools for image analysis and adding structure to the Wikimedia Commons repository. 
Tools built by community members also support editors in discovering and adding images in Wikimedia projects, including the WDFIST~\cite{wdfist} and the Wikishootme~\cite{wikishootme} tools. 
Finally, several community initiatives are organized around adding pictorial representations of under-represented people and topics, such as Wiki Loves Monuments~\cite{wikilovesmonuments}, Visible Wiki Women~\cite{visiblewikiwomen}, and Wikipedia Pages Wanting Photos~\cite{wpwp}.
\subsubsection{Structured Data}
The \textit{structured data} gap reflects the differences in the use of structured data across Wikimedia projects, namely the amount of information that is organized and indexed in a machine-readable fashion. Structured data offers the potential for managing data for Wikimedia projects on a global scale, allowing easier indexing and offering tools for multilingual data creation and description. One of the most prominent examples of structured data is Wikidata, a structured data knowledge base which was designed with the goal of increasing knowledge diversity~\cite{Vrandecic2014wikidata}. More structured data is not a panacea for equity, however, and greater reliance on structured data can concentrate power and remove important context~\cite{ford2016provenance,carroll2020care}.
Information from Wikidata is being re-used in Wikipedia articles in different ways, most notably in infoboxes~\cite{wpwikidata} but perhaps most commonly in metadata templates~\cite{wikidatareuse}. 
The Wikidata Concepts Monitor~\cite{wdcm_usage} provides quantitative insight into the degree of Wikidata usage, albeit inexactness in tracking this usage makes it non-trivial to interpret the statistics with respect to individual templates or infoboxes.
In addition, it must be noted that, beyond some established cases, the degree to which Wikidata should be used in Wikimedia projects is part of ongoing discussions within the respective communities (see for example the case in English Wikipedia~\cite{useofwikidata}).
Given the importance and potential of structured data, the Wikimedia Foundation's Medium Term Plan 2019-2020 (Platform Evolution) aims at increasing its usage across Wikimedia projects~\cite{mtp}. Product initiatives such as the Structured Data on Commons program~\cite{sdc}, as well as new proposals for an Abstract Wikipedia~\cite{vrandecic2020architecture} aim at closing this gap and enable communities to consume and contribute structured data across languages.
Along with structured data on Wikimedia projects, there are many semi-structured annotations that play an important role in the maintenance and curation of Wikimedia content. For example, the category network helps with content discoverability and classification, various extensions such as PageAssessments~\cite{pageassessments} can be used to rate the quality and importance of articles to help in evaluating and prioritizing content, numerous templates are used for tracking issues to be resolved in content~\cite{maintenancetemplates}. While these annotations are often hidden to readers, they are crucial to research and tools aimed at discovering and improving content.
\subsubsection{Readability}
The \textit{readability} gap refers to how difficult it is to read the content of an article or a piece of information in Wikimedia projects in comparison to the reading abilities of their readers.
Using statistical tools such as the Gunning-Fog index,\footnote{\url{https://en.wikipedia.org/wiki/Gunning_fog_index}} researchers have tried to quantify this gap by computing language complexity of texts in Wikipedia articles~\cite{Yasseri2012language}.
Different studies~\cite{lucassen2012readability,brezar12019readability} have shown that the difficulty is considerably above the reading ability of average adults even for projects such as Simple English Wikipedia, which are explicitly aimed at closing this gap.

An indirect approach to bridge the gap is to make content available in different languages given the high proportion of non-native readers in some languages~\cite{2019survey}. 
For this, researchers developed machine-learning models to discover and prioritize articles missing in a given Wikipedia language edition which, in turn, are recommended for creation to editors~\cite{wulczyn2016growing}. 
Platforms such as the Content Translation Tool~\cite{contenttranslation} help in translating articles to any language available in Wikipedia.

\subsection{Barriers}\label{sec:contentbarriers}

Wikipedia content is governed by three principal core content policies---Neutral point of view, Verifiability, and No original research---which define the scope and the material that should exist in the online encyclopedia.
These policies shape the way in which content is added to Wikipedia, and, to some extent, to its sister projects. While these policies ensure that the content in the project is of high quality, they have also been identified as major barriers for diverse content inclusion~\cite{genderequity2018}.  
\subsubsection{Verifiability}
Wikipedia's Verifiability core content policy~\cite{verifiability} requires every piece of information that has been challenged---or is likely to be challenged---to be backed by a reliable source. The accessibility of a source and its availability in a specific language play a key role in the coverage of a piece of content in different projects~\cite{sen2015barriers}.
Previous research has deepened our understanding of citation usage by highlighting gaps in terms of the types of sources used~\cite{Ford2013,sen2015barriers}, barriers to inclusion of relevant sources around content about marginalized communities~\cite{2020readingtogether}, or their accessibility~\cite{Harder2015} as well as in developing methods for how to automatically detect the citation span of individual sources~\cite{Fetahu2017} or classify what content is missing citations~\cite{Redi2019citationneeded}. Censorship outside of Wikimedia---i.e. more general political freedom within a country as measured by organizations like Reporters Without Borders or Freedom House---can also have a large impact on the availability of reliable sources and balanced coverage of topics.
There exist different initiatives to help editors monitor citation quality at scale such as the Citation Detective tool~\cite{choucitation}, which automatically produces dumps of sentences requiring citations in English Wikipedia, or Wikicite~\cite{wikicite}, whose goals include the ``improvement of citations in Wikimedia projects and an open, collaborative repository of bibliographic data for innovative applications''. 
Furthermore, the Wikipedia Library portal~\cite{wikipedialibrary} constitutes a fundamental tool to help breaking the verifiability barrier, as it provides reliable sources that editors can use to improve the articles, as well as promoting community initiatives such as the ``1Lib1Ref'' campaign, which encourages librarians to add references missing in Wikipedia. Furthermore, tools like Scholia~\cite{NielsenF2017Scholia}, a service that creates visual scholarly profiles for different items such as topics, people, and organizations using bibliographic and other information in Wikidata, can help retrieve specific sources needed to support a claim.
While citations and references are a key tool to monitor the quality of knowledge coming from written sources, how to incorporate oral knowledge within the verifiability framework, remains an open point for discussion~\cite{oralknowledge}.

\subsubsection{Neutrality}
The Neutral Point of View (NPOV) policy~\cite{npov} demands that all encyclopedic content should represent all significant views fairly, proportionately, and without bias. 
To help in understanding and bridging this gap, researchers in various fields have tried to characterize the dynamics of NPOV disputes~\cite{hassine2005dynamics,pavalanathan2018mind}, as well as quantify the NPOV gap from explicit (statements supporting a certain POV) or implicit (omission of certain aspects) bias~\cite{Hube2017bias}. Examples include the use of NPOV-templates to automatically detect biased language in the content of Wikipedia~\cite{Recasens2013,Hube2018}, or documenting biases in specific topics such as politics~\cite{Greenstein2012bias} or culture~\cite{Callahan2011culture} among others.
The Neutral point of view Noticeboard~\cite{npovnoticeboard} is one of the main sub-communities dedicated to the discussion of neutrality of content in English Wikipedia. Similar to the verifiability case, incorporating non-written knowledge which complies with the NPOV policy is still under discussion~\cite{oralknowledge}.

\subsubsection{Notability}
Wikipedia contains several other policies that constitute potential barriers for content, most prominently the guidelines around notability~\cite{notability}. The criterion of notability is used to decide whether a topics warrants its own article on Wikipedia. In its most general form it requires that ``a topic has received significant coverage in reliable sources that are independent of the subject'' for a topic to warrant its own article.
The interpretation of these guidelines is highly controversial (see, for example, the debate on deletionism vs inclusionism ~\cite{deletionism}). Notability has been claimed to play a key role in causing and amplifying systemic biases in Wikipedia~\cite{systemicbias}. For example, it is hypothesized that notability is harder to establish for certain topics due to lack of English sources. In fact, research on discussions for deletions~\cite{lam2009wikipedia,schneider2012deletion} has shown that notability is the most prominent indicator for deletion of articles. Case studies, such as on the deletion of biographies of female scientists, have received wide media attention.
For the example of gender bias, several studies~\cite{Wagner2016glass,Schellekens2019recognition} have shown that female biographies on Wikipedia are more notable than men using external signals of notability such as Google Search volume.
However, overall notability remains an understudied area with respect to its role in content gaps.

\subsubsection{Free Knowledge}
A major barrier to providing diverse, high-quality content is the availability and openness of knowledge. For example, laws related to the freedom of panorama, which allows people to freely take photos of otherwise copyrighted work such as public art, varies greatly country to country.\footnote{\url{https://en.wikipedia.org/wiki/Freedom_of_panorama}} Strict rules around what licenses are permitted for use on Wikimedia Commons creates clarity for end-users~\cite{commonsreusers} but restricts what content can be uploaded---e.g., valuable health-related content from the World Health Organization.\footnote{\url{https://www.who.int/about/who-we-are/publishing-policies/copyright}} These restrictions reduce the availability of content and also can be challenging to understand and navigate for newcomers~\cite{glamcommons}.

%% file: tables/content_table.tex
\begin{table}[]
\resizebox{\linewidth}{!}{
\renewcommand{\arraystretch}{1.2}
\begin{tabular}{p{3.2cm}|p{2.5cm}p{5.5cm}p{4.8cm}}
  \textit{\textbf{Facet}} &
  \textit{\textbf{Gap}} &  \textit{\textbf{Description}} &
  \textit{\textbf{Sources}}\\
  \hline
\multirow{5}{3.2cm}{\textbf{~{ }\\Representation }\\~{ }
}
&&&\\
  &
    \textit{Gender} 
&
   Differences in content coverage  depending on the gender identity of subjects &
  literature~\cite{Reagle2011britannica,Bamman2014unsupervised,Eom2015interactions,graells2015first,Wagner2015mans,Wagner2016glass,Young2016ceo,Halfaker2017keilana,Menking2017who,Zagovora2017professions, konieczny2018gender,Adams2019who,Hinnosaar2019newmedia,Schellekens2019recognition}, strategy~\cite{topictrategy,mtp}, 
  community~\cite{counteringsystemicbias,whoseknowledge,whgi,denelezh,wdcm_bias,gendergapportal,wikiprojectwomen,visiblewikiwomen}
   \\
 &
  \textit{Age (Recency)} &
  Differences in coverage of topics and facts related to different times in history &
  literature~\cite{samoilenko2017analysing}, 
  community~\cite{wikiprojecthistorical}
  \\
 &
  \textit{Geography} &
   Differences in coverage of topics related to geographic regions or population distribution  &
  literature~\cite{Hecht2009selffocus,Graham2011geographies,Graham2014geo,Beytia2020positioning}, strategy~\cite{topictrategy,mtp}, 
  community~\cite{counteringsystemicbias,whoseknowledge,africaportal,wikilovesafrica}
   \\
    &
  \textit{Language} &
  Differences in coverage of topics and facts depending on the content language &
  literature~\cite{wulczyn2016growing,he2018the_tower_of_babel,bao2012omnipedia}, 
  community~\cite{sdc}
  \\
   &
  \textit{Socioeconomic Status} &
  Differences in coverage of topics and facts related to different socioeconomic contexts &
  literature~\cite{miquel2019wikipedia,rask2008reach}, 
  strategy~\cite{},
  community~\cite{counteringsystemicbias}
  \\
   &
  \textit{Sexual Orientation} &
  Differences in coverage of topics and facts related to different sexual orientations &
  literature~\cite{park2020multilingual}, 
  community~\cite{wikidatalgbt}
  \\
 &
  \textit{Important topics} &
  Differences in coverage of topics that are of common interest &
  literature~\cite{Kittur2009whatsinwikipedia,Lewoniewski2019multilingual,Shafee2017evolution,Smith2020health},
  strategy~\cite{topictrategy,mtp}, 
  community~\cite{counteringsystemicbias,whoseknowledge,vitalarticles,wikiprojectmedicine,allhumanknowledge}
   \\
 &
  \textit{Cultural Background} &
  Differences in coverage of topics related to the history, heritage, and characteristics of a current or former cultural group &
  literature~\cite{MiquelRibe2016,MiquelRibe2018}, 
  strategy~\cite{topictrategy,mtp},
  community~\cite{counteringsystemicbias,whoseknowledge,wcdo,frenchcarribeanculture}
  \\
  \hline
\multirow{4}{3.2cm}{\textbf{~{ }\\Interaction }\\~{ }
}
&&&\\
  &\textit{Multimedia} &
  Differences in coverage with respect to the type of media used to share the content &
  literature~\cite{Viegas2007,He2018babel,Porter2020visual}, 
  strategy~\cite{strategy,innovationstrategy}, 
  community~\cite{imageclassifiers,imagerole,imagewikidata,imagewikipedia,imageusepolicy,wdfist,wikishootme,wikilovesmonuments,visiblewikiwomen}
   \\
  &\textit{Structured Data} &
Differences in the use of information which is indexed and machine-readable &
  literature~\cite{Vrandecic2014wikidata, vrandecic2020architecture}, strategy~\cite{strategy,mtp}, community~\cite{wpwikidata,wdcm_usage,useofwikidata,sdc}
 \\
 &
  \textit{Readability} &
Differences in the ease with which textual content can be understood by readers of different backgrounds &
  literature~\cite{Yasseri2012language,lucassen2012readability,wulczyn2016growing,brezar12019readability}, 
  strategy~\cite{strategy},
  community~\cite{contenttranslation,manualofstyle,dosanddonts}
  \\
  \\\hline
\end{tabular}}
\end{table}

%% file: sections/5_How_to.tex
\section{How to Use this Taxonomy}\label{sec:how-to}

The taxonomy of knowledge gaps developed in this research and the corresponding literature review can be utilized in a variety of ways by different actors in the free knowledge ecosystem and the Wikimedia projects. This includes Wikimedia community organizers, affiliates, Wikimedia Foundation staff, contributors, researchers and partners. Below we describe the different aspect of the taxonomy and how they can be utilized.

\noindent
\textbf{The taxonomy as a framework for conversations.} As evidenced through the work in developing the taxonomy of knowledge gaps, Wikimedia projects face many different knowledge gaps. However, prior to the development of this taxonomy, the bulk of the attention of large actors within the Movement such as the Wikimedia Foundation has been heavily focused on specific gap types. We hope that the more comprehensive taxonomy of gaps can provide a framework for the decision makers to learn about the different gap types, brainstorm about their possible relationships, and devise ways to address them. 

\noindent
\textbf{The references that support the gaps.} As part of the process of developing the taxonomy of knowledge gaps, we conducted a major literature review. This review of past work which is presented as \textit{sources} in the tables throughout this paper as well as in the reference section can be a valuable resource for those who are eager to expand their knowledge of the particular gaps and learn about their possible causes.

\noindent
\textbf{A first step towards measuring knowledge gaps.} The definitions of the different gap types provided through the taxonomy are a necessary step for developing metrics to measure the different gap types. The measurement of the gap types and understanding the relationships between them can further help to develop the \textit{knowledge gap index}, a composite index that can operationalize knowledge equity in the Wikimedia projects and help support more data-informed decision-making processes.

%% file: sections/6_Future.tex
\section{Future Work}\label{sec:future}
This taxonomy represents a first step towards understanding the underlying mechanisms that prevent us from reaching knowledge equity, and designing solutions to remove those barriers preventing people from accessing free knowledge. In this Section, we identify the next steps to reach this goal by building on top of the taxonomy presented in this manuscript.
\subsection{Metrics}
In order to track progress towards knowledge equity, it is necessary to quantify each of the knowledge gaps described in this taxonomy.  
For this we need to i) carefully define metrics reflecting the gap extent across different categories, and, ii) identify relevant reference data sources to contextualize the metric and measure the amplitude of the gap. The main challenges in this endeavour are:
 \begin{itemize}
    \item Operationalizing a gap: The definition of each gap must be translated into a set of measurable quantities.
    For example, for the content gender gap in Wikipedia, a common choice is to use the property P21 (sex or gender)\footnote{\url{https://www.wikidata.org/wiki/Property:P21}} of the corresponding item in Wikidata. While that approach is simple, transparent, and has good data coverage, questions can still be raised about the appropriate dimensions to report and how to capture articles that are not explicitly about a person (such as women's health). Operationalizing other gaps might be even less straightforward.
    The case of topical content gaps serves as an illustrative example. Identifying content related to, e.g., medicine, is possible through manual annotation in Wikiprojects~\cite{wikiprojectmedicine} though it is time-intensive and requires constant updating which might be unfeasible for the many small and medium-sized projects. Automatic methods such as the ORES topic classification~\cite{ores} alleviate some of these issues, however, they are currently limited in their general applicability as they only support a small set of languages and a fixed set of 64 topics.
    
     \item Availability of data: Some data for specific gaps is just not available or hard to obtain for privacy or other reasons. For example, obtaining information about readers' ages in order to measure the age representation gap requires careful legal frameworks~\cite{2019survey}. Succinctly capturing complex and contextual social phenomena like race is largely not possible at a global scale.
     
     \item Consistency of data: seasonality, large-scale events or holidays, and edit campaigns can have a large impact on who is reading or contributing to Wikipedia at any given moment in a year~\cite{chelsy2019detecting}. This becomes particularly salient when trying to survey a global population where the effects of these events will vary greatly between regions and even hemispheres of the world. These external effects will have to be taken into account for data that is collected at irregular (or even regular) intervals. On top of this, varying sampling methodologies or survey questions can also greatly impact who responds and how~\cite{howmanywomeneditwikipedia}.
     
     \item One gap, many metrics: For every individual gap, there exist multiple, equally relevant, ways in which to measure it. Thus, it is important to note that there are many metrics which capture different aspects of the same gap. The content gender gap provides an illustrative example, as it constitutes one of the most well-studied gaps. Public tools documenting the gap (such as Wikidata Human Gender Indicators~\cite{whgi} or Denelezh~\cite{denelezh}) document the number of articles on people of different gender identities, respectively, thus capturing only the selection of content. Research has shown that it is equally important to consider other aspects such as extent (e.g., comparing the quality of articles on men and women~\cite{Halfaker2017keilana}) or the framing (e.g., comparing the language in the articles of men and women~\cite{Wagner2016glass}). Building on initial insights~\cite{contentgapswikipedia}, future research needs to understand the role of different metrics of the same gap and how the well-explored content gender gap might inform development of metrics for less-explored facets.

     \item Goal: The definition of a metric for a gap allows one to make a statement about whether the gap is small or large. Such definition thus leads to conclusions about whether a gap is closed or not. What this goal should be might not be as obvious as it seems. For example, in the context of the content gender gap, recent discussions in the community revolve around what are suitable baselines for the comparison of the number of biographies on men and women~\cite{sizegendergapactually}. Future work involves curation of external datasets such as sociodemographic data on country-level and consultation and discussions with communities on how to define the metrics.  
 \end{itemize}
\subsection{Knowledge Gap Index}
One of the end goals of this taxonomy is to generate a knowledge gaps index, namely a single indicator combining metrics from different gaps, measuring the overall knowledge equity of Wikimedia projects. Socioeconomic indices have been adopted by many organizations, advocacy groups and policy makers such as the Global Innovation Index,\footnote{\url{https://www.globalinnovationindex.org/Home}} the Human Development Index from the United Nations,\footnote{\url{http://hdr.undp.org/en/content/human-development-index-hdi}} or the Gender Equality Index from the European Union.\footnote{\url{https://eige.europa.eu/gender-equality-index/}}
Such indices are a useful tool for analysis of policies and communication with the public because they can give a coarse-grained view on complex multi-dimensional realities.
Indices require the projection and aggregation of different metrics into a composite index and are developed over the course of several steps, the first being the construction of a theoretical framework~\cite{Commission2008handbook}. 
The main challenges to reach this goal are:
\begin{itemize}
    \item Availability of metrics: An index requires the availability of validated metrics (see the discussion in the previous subsection).
    \item Robustness: Even if all metrics are available, the robust construction of a composite index is non-trivial. According to recommended procedures, this taxonomy indeed represents only the first, theoretical step towards a comprehensive, robust and high-quality index~\cite{Commission2008handbook}. Additional steps would include, for example, uncertainty and sensitivity analysis.
\end{itemize}

\subsection{Interactions between Gaps}
The taxonomy presented here reflects a simplified version of the complex landscape of knowledge gaps by depicting readership, contributorship, and content as three independent dimensions. However, these dimensions are interrelated and feed into each other, as pointed out by the community~\cite{systemicbias}, research~\cite{shaw2018pipeline}, and Wikimedia Foundation in their medium-term plan~\cite{mtp}.
Apart from few examples (e.g.~\cite{WarnckeWang2015misalignment} on the misalignment between content and readership), our understanding of how different gaps interact with each other is still very limited, and part of our future research work includes shedding light on these interactions.

\subsection{Causes and Interventions}
In order to bridge knowledge gaps, we not only need metrics to monitor the efficacy of individual interventions, but also require an understanding of the underlying causes to design effective interventions. 
However, causal evidence on what drives gaps is scarce. Recent research has identified several candidate hypothesis for individual gaps~\cite{explainingreadergendergap}.
An example comes from previous work analyzing knowledge gaps in readership and contributorship~\cite{shaw2018pipeline}: the authors found that the gender gap is higher in contributors than in readers, and proposed interventions including continued efforts in female editors recruitment, as well as awareness campaigns to increase awareness that Wikipedia is editable among women readers.

The existence of power asymmetries in society\footnote{For further reading, see~\cite{d2020data}} are central to this taxonomy, knowledge equity~\cite{strategy}, and whether any particular facet is included. Wikimedia reflects the broader world, therefore systematic inequalities in power lead to encyclopedic knowledge being systematically biased against specific communities~\cite{systemicbias,ford2017anyone,whoseknowledge}. 
Some power asymmetries are more consistent globally---e.g., gender, urban versus rural, socioeconomic status---while others are much more contextual and regional---e.g., race, religion. The consistency of a power asymmetry does not make a gap more or less important, but it does affect the ability to characterize and quantify a gap and thus its relevance to the guiding principles of this taxonomy (Section~\ref{sec:guidingprinciples}). 
Addressing the gaps listed in this manuscript is not therefore simply a matter of creating more content or bringing in a more diverse community. Analogous to the deep body of work discussing the relationship between societal inequalities and bias within AI systems~\cite{west2019discriminating,barocas2017fairness,blodgett-etal-2020-language,havens2020situated}, addressing these gaps in the long-term will also require addressing these societal inequalities, along with the many, more-Wikimedia-specific barriers that drive these gaps---see Sections~\ref{sec:readerbarriers}, \ref{sec:contribbarriers}, and \ref{sec:contentbarriers}.

\subsection{Expanding the Taxonomy}
The taxonomy itself is also incomplete (and will always be) but it is worth detailing a few salient gaps in the process itself:
\begin{itemize}
    \item Positionality: while we, the authors, sought to build on as wide of a range of sources and incorporated feedback from many in the community~\footnote{\url{https://meta.wikimedia.org/wiki/Research_talk:Knowledge_Gaps_Index/Taxonomy}}, we represent a narrow slice of backgrounds and language communities. Notably, we generally are able-bodied, employed by the Wikimedia Foundation, and share an academic computer science background from institutions in the United States and Europe. Our common language is English. As such, we almost certainly missed or misrepresented aspects that others with different backgrounds would have incorporated differently. Addressing these gaps will require decentering our backgrounds and centering marginalized knowledge~\cite{centeringmarginalized}.
    \item Sources: while we sought to gather sources from many areas, we almost certainly have blind spots in our literature review due to our positionality---e.g., fields of study such as science and technology studies (STS), languages beyond English and the others we speak. Additionally, the research community as a whole has largely focused its study on Wikipedia, with a growing literature about Wikidata and Wikimedia Commons, but the research is scant about Wiktionary, Wikisource, and other projects.
    \item Types of knowledge: while this taxonomy seeks to cover the Wikimedia projects, there are certainly types of knowledge that sit outside of the scope of the existing Wikimedia projects and therefore are not taken into consideration here. For example, Wikifunctions~\cite{abstractwikipedia} was established as the newest Wikimedia project in 2020 and there are certainly other forms of knowledge that are not covered by the existing projects -- e.g., Jimmy Wales' list of things that need to be free~\cite{thingsneedfree}.
    \item Other types of contributors: the contributors gap (Section~\ref{sec:contributors}) is specifically scoped to people who are editing the Wikimedia projects. There are many other ways to contribute to Wikimedia though outside of editing content. Most notably not included in this definition are organizers and developers. Organizers are central to the success of the Movement and advancing equity in participation and content through work such as campaigns or advocacy~\cite{organizerstudy}---for more information, see the Organizer Framework~\cite{organizerframework} and survey data from Community Insights~\cite{2020communityinsights}. Developers build the Mediawiki infrastructure, associated gadgets, templates, and bots that allow for editing of Wikimedia projects. This community has a large impact on equity in the projects both directly---e.g., bot-generated articles---and indirectly---e.g., by what power relations they encode in the infrastructure such as acceptable or default genders~\cite{ford2017anyone}. Survey data about different parts of this community can be found in the Community Insights surveys~\cite{2020communityinsights}, Wikimedia Foundation diversity report~\cite{wmfdiversity}, and Cloud Services Annual Survey~\cite{developersurvey}.
    
\end{itemize}

%% file: sections/7_Acknowledgements.tex
\section{Acknowledgments}
We would like to thank all of the affiliates, user groups, and volunteer community members as well as the Wikimedia Foundation staff who engaged in providing feedback to the first draft of the taxonomy through surveys, meetings, and the Wikimedia page for the project\footnote{\url{https://meta.wikimedia.org/wiki/Research_talk:Knowledge_Gaps_Index/Taxonomy}}. Thanks to these contributions, this second version of the Taxonomy of Knowledge gaps is much more solid and aligned with the spirit and strategic directions of the Wikimedia community.